\documentclass[prx,superscriptaddress,twocolumn,noshowpacs]{revtex4-1}
\usepackage[pdftex]{graphicx}
\usepackage{amsmath}
\usepackage{amssymb}
\usepackage{epstopdf}
\usepackage{tikz}

\usepackage{times}
\usepackage{enumitem}
\usepackage{comment}
\usepackage{subcaption}
\usepackage{braket}
\usepackage{dsfont}

\newcommand{\paul}[1]{{\color{blue}\footnotesize{(PF) #1}}}

\DeclareMathOperator{\Tr}{Tr}

\newcommand{\superscript}[1]{\ensuremath{^{\textrm{#1}}}}

\renewcommand{\th}[0]{\superscript{th}\,}

\newcommand{\centerfloat}{%
  \parindent \z@
  \leftskip \z@ \@plus 1fil \@minus \textwidth
  \rightskip\leftskip
  \parfillskip \z@skip}

\tikzset{dangerous style/.code={
    \tikzoption{clip}[]{\pgf@relevantforpicturesizefalse}
    \tikzoption{use as bounding box}[]{\pgf@relevantforpicturesizefalse}
    }
}

 \definecolor{blue}{HTML}{00538A} 
  \definecolor{red}{HTML}{C10020} 
  \definecolor{SkyBlue}{HTML}{CEA262}
\definecolor{orange}{HTML}{FF6800}

\begin{document}

\title{Long coherence times for edge spins}
\author{Jack Kemp}
\affiliation{Rudolf Peierls Centre for Theoretical Physics, University of Oxford, 1 Keble Road, Oxford, OX1 3NP, UK}
\author{Norman Y. Yao}
\affiliation{Department of Physics, University of California, Berkeley, CA 94720, USA}
\author{Christopher R. Laumann}
\affiliation{Department of Physics, Boston University, Boston, MA, 02215, USA}
\author{Paul Fendley}
\affiliation{Rudolf Peierls Centre for Theoretical Physics, University of Oxford, 1 Keble Road, Oxford, OX1 3NP, UK}
\affiliation{All Souls College, University of Oxford, Oxford OX1 4AL UK}


\begin{abstract}

We show that in certain one-dimensional spin chains with open boundary conditions, the edge spins retain memory of their initial state for very long times. The long coherence times do not require disorder, only an ordered phase. In the integrable Ising and XYZ chains, the presence of a strong zero mode means the coherence time is infinite, even at infinite temperature. When Ising is perturbed by interactions breaking the integrability, the coherence time remains exponentially long in the perturbing couplings. We show that this is a consequence of an edge ``almost'' strong zero mode that almost commutes with the Hamiltonian. We compute this operator explicitly, allowing us to estimate accurately the plateau value of edge spin autocorrelator.

\end{abstract}

\maketitle
\section{Introduction}

A fundamental desire in quantum engineering is for localized degrees of freedom to maintain coherence over long times. While of course many of the issues in achieving such are experimental, recent theoretical work has provided a variety of new avenues to explore. Such avenues include  1) finding systems exhibiting prethermalization, where achieving thermal equilibrium takes unusually long, 2)  exploiting topological invariants, and 3) adding strong disorder so that the ensuing many-body localization guarantees the existence of many conservation laws. 

In this paper we analyze a type of quantum coherence involving aspects of all three. We show that in ordered phases in certain quantum spin chains, an almost {\rm edge strong zero mode} operator results in unusually long coherence times for the edge spins. This behavior is very similar to what happens in prethermalization \cite{Essler:2014,Abanin}. Such zero modes are familiar from the study of topologically ordered systems \cite{KitMajorana}, and can guarantee degeneracies in the spectrum, even among highly excited states \cite{Fendley:2016}. Unusual behavior in highly excited states is a hallmark of many-body localization \cite{Husereview}, and a related phenomenon is described in \cite{HNOPS,Bauer,Bahri}. Similar behavior also occurs in the integrable XXZ chain, at least in excited states with zero energy density \cite{Masud1,Masud2}.  

We show here that a long edge-spin coherence time requires neither disorder and nor integrability. In fact, we show that if integrable, an edge strong zero mode guarantees that the edge-spin coherence time is infinite, up to finite-size effects. In the presence of integrability-breaking interactions, the oprerator becomes an ``almost'' edge strong zero mode, whose presence makes this coherence time finite but very long.  This operator is quite reminiscent of, and presumably related to, the slowly relaxing local operators found by a numerical search \cite{Kim:2015}. Here however we are able to construct the zero mode directly and so analyze how it changes as couplings are varied. With one type of perturbation, we show that the coherence time decreases with the strength of the integrability breaking, as one might expect. However, when a different ordering term is included in the Hamiltonian, the coherence time is not monotonic in its strength. Rather, it depends on some interesting physics also seen in the many-body-localized context, typically known as resonances \cite{Ros:2014}. One fascinating consequence is that the coherence time is independent of whether or not the perturbation favors a competing order; the physics here is governed mainly by excited-state properties oblivious to the precise form of order in the ground state.

In section II, we explain how in the Ising and XYZ spin chains, a strong edge zero mode results in an infinitely long
coherence time for the edge spin. Only finite-size effects, exponentially small, destroy the coherence. Section III is the core of the paper. Here we show how in a non-integrable model, the Ising chain with additional interactions, the edge spin has a finite but very long coherence time. The reason is the presence of an ``almost'' strong zero mode, which almost commutes with the Hamiltonian. We compute it explicitly, and use it to estimate the value of the edge spin autocorrelator before the decay. In section IV, we address one reason why the coherence time is finite, the presence of resonances in perturbation theory. These turn into poles in the expansion of the strong zero mode. Section V contains some conclusions.

\section{Infinite coherence time from the strong edge zero mode}

\subsection{The basic idea}
\label{sec:basic}

We study quantum spin chains with $L$ sites and open boundary conditions. For simplicity, we focus on systems with two states per site. A convenient basis for operators acting on the $2^L$-dimensional Hilbert space is given by products of $\sigma^a_j$,  the Pauli matrix $\sigma^a$ acting at site $j$ and trivially on the others, i.e.\ 
$\sigma^a_j = 1\otimes 1\otimes\dots 1\otimes \sigma^a \otimes 1 \otimes\dots \otimes 1$. 
We study models with a ${\mathbb Z}_2$ symmetry under flipping all spins, where the operator
\begin{align}
{\cal F} = \prod_{j=1}^L \sigma^x_j
\end{align}
commutes with the Hamiltonian. We focus on ordered phases where
the spin-flip symmetry is spontaneously broken, i.e.\ in infinite volume the system has two distinct ground states $|g_+\rangle$ and $|g_-\rangle$ with ${\cal F}|g_\pm\rangle =\pm|g_\pm\rangle$. An important example is the transverse-field Ising chain, with Hamiltonian
\begin{equation}
H_{\rm Ising} = -J\sum_{j=1}^{L-1} \sigma^z_j  \sigma^z_{j+1} - \Gamma \sum_{j=1}^L \sigma^x_j\ .
\label{HIsing}
\end{equation}
The ordered phase occurs for  $|\Gamma|<|J|$.

Although we choose couplings so that the ground state is ordered, our interest is in the behavior of highly excited states. 
The basic physical quantity we study is the autocorrelator of the edge spin magnetization 
\begin{align}
{A}_s(t) \equiv \Braket{s|\sigma^z_1(t)\sigma^z_1(0)|s}\ ,
\end{align}
where $s$ is an eigenstate of the Hamiltonian.
In an ordered phase, the boundary magnetization in the ground states $\Braket{g_\pm|\sigma^z_1|g_\mp} $ is non-vanishing for open boundary conditions. Thus not surprisingly, $A_{g_\pm}(t)$ is non-vanishing as $t\to\infty$. Remarkably, we will show that in certain integrable systems such an infinite coherence time holds for {\em all} eigenstates of $H$:
\begin{align}
\lim_{t\to\infty} \lim_{L\to\infty}A_s(t) \ne 0\ .
\label{Alim}
\end{align}
Our central result is an even more profound statement: even when integrability is broken and $s$ is a highly excited eigenstate (or some mixture of them), ${A}_s(t)$ can decay very slowly with time.  This slow decay is very reminiscent of ``prethermalization'' in systems after a quantum quench \cite{Essler:2014,Abanin,Kim:2015}.

To understand this slow decay, we first explain why the infinite edge-spin coherence time (\ref{Alim}) occurs in certain integrable spin chains. It is because a {\em strong zero mode} results in a ``pairing'' in the spectrum \cite{Fendley:2012,Jermyn:2014,Alicea:2015,Fendley:2016,Muller:2016}. The connection between the pairing and the infinite coherence time becomes apparent by introducing a resolution of the identity into the autocorrelator:
\begin{align}
\nonumber 
{A}_{s}(t)  &= \sum_r \Braket{s|e^{-i H t}\sigma^z_1e^{i H t}| r}\Braket{r|\sigma^z_1|s} \\
&=  \sum_r |\Braket{s |\sigma^z_1| r}|^2 e^{i\left(E_{r}-E_s\right)t}
\label{Asum}
\end{align}
where $\ket{s}$ and the $\ket{r}$ are eigenstates of $H$ with energies $E_s$ and $E_r$ respectively, and $\hbar=1$.  Because $\{\sigma^z_1,{\cal F}\}=0$, we can restrict the sum over $r$ to those states obeying ${\cal F}|r\rangle =-{\cal F}|s\rangle$. 
When $|s\rangle$ is a highly excited eigenstate, typically the matrix element $\Braket{s |\sigma^z_1| r}$ is non-vanishing and small for many states $r$ with many different energies $E_r$. The sum in (\ref{Asum}) then contains many incoherent oscillating factors making $A_s(t)$ decay very rapidly in time.  However, if there is a state $r$ where both 
\begin{enumerate}
\item  $\Braket{r|\sigma^z_1|s}$ is finite,
\item $E_r\approx E_s$
\end{enumerate}
then it immediately follows from (\ref{Asum}) that the edge coherence time is infinite in the large-size limit, as in (\ref{Alim}).
By $E_r\approx E_s$ we mean up to corrections exponentially small in system size.

We will explain precisely the consequences of a strong zero mode in the next subsection \ref{sec:strong}. 
Here we illustrate the pairing in a simple example, the special case $\Gamma=0$ of the Ising chain familiar from the studies of topological order \cite{KitMajorana}. The eigenstates and eigenvalues of the Hamiltonian $H_0$ with $\Gamma=0$ are given by specifying all the eigenvalues of the $\sigma^z_j$. The states $|+++\dots ++\rangle$ and $|---\,\cdots\,--\rangle$ are both ground states of $H_0$ when $J>0$. They are not eigenstates of spin-flip symmetry, but these are easily found:
\begin{align*}
\ket{g_\pm}&= \frac{1}{\sqrt{2}}(|+++\dots ++\rangle\pm |---\dots--\rangle)\\
&= \frac{1\pm {\cal F}}{\sqrt{2}}|+++\dots ++\rangle
\end{align*}
All eigenstates of both $H_0$ and ${\cal F}$ can be written in the form
\[\ket{s_\pm}= \frac{1\pm {\cal F}}{\sqrt{2}} |+\pm\pm\pm\dots \rangle\]
for all $2^{L-1}$ choices of the $\pm$ signs. Since $[{\cal F},H]=0$ and ${\cal F}^2=\mathds{1}$, the energies obey $E_{s_+}=E_{s_-}$. One way of toggling between these degenerate states is simply to measure the edge spin. Because $\sigma_1^z$ anticommutes with ${\cal F}$, 
\[\Braket{s_\pm|\sigma^z_1|s_\mp}=1\ .\]
Thus in this limit, $A_{s_\pm}(t)=1$ for all $t$ and $s_\pm$. This of course is not surprising, given there are no off-diagonal terms in the Hamiltonian. What is remarkable is that not only does this pairing persist for $\Gamma$ non-vanishing, but that it also occurs in an interacting model, the XYZ spin chain.

\subsection{The strong zero mode}
\label{sec:strong}

The edge-spin coherence (\ref{Alim}) arises as a consequence of a {\em strong zero mode} \cite{Fendley:2012,Jermyn:2014,Alicea:2015,Fendley:2016}. 
A strong zero mode is an operator $\Psi$ that maps an eigenstate in one symmetry sector to that in another with the same energy up to exponentially small finite-size corrections. Precisely, in our examples $\Psi$
\begin{enumerate}
\item almost commutes with the Hamiltonian: $[H,\Psi] = {\cal E}$, where the ``error'' ${\cal E}$ is an operator with norm  $|{\cal E}|<e^{-\alpha L}$ with $\alpha$ a positive constant,
\item anticommutes with spin-flip symmetry:   $\{{\cal F}, \Psi\} = 0$,
\item squares to the identity operator: $\Psi^2 = \mathds{1}$.
\end{enumerate} 
Because $[{\cal F},H]=0$, eigenstates of $H$ can be organized into sectors with ${\cal F}=\pm 1$. The presence of a strong zero mode guarantees that the {\em entire} spectrum in the ${\cal F}=1$ sector is the same as that in the ${\cal F}=-1$ sector, up to corrections of order $e^{-\alpha L}$. Namely, eigenstates of $H$ form ``pairs'' $|s_+\rangle$ and $|s_-\rangle$, with $E_{s_+}\approx E_{s_-}$ and $|s_\pm\rangle \approx \Psi |s_\mp\rangle$. Here and henceforth $\approx$ means equal up to corrections with norm vanishing exponentially fast as $L\to\infty$. 

The most famous example of a strong edge zero mode is that localised on the edge of the Ising chain \cite{Pfeuty,KitMajorana}.  We showed at the end of section \ref{sec:basic} that when $\Gamma=0$, the operator $\sigma^z_1$ obeys all three conditions above, and so pairs each eigenstate in the ${\cal F}=1$ sector with one with ${\cal F}=-1$ and the same energy. The fermionic version of the Ising Hamiltonian (\ref{HIsing}) is often known as the Kitaev chain, with the ${\mathbb Z}_2$ spin order corresponding to topological order. The operator $\sigma^z_1$ is the Majorana fermion operator on the edge, which does not appear in $H_0$ and so commutes with it. In phases with topological order, edge zero modes mapping the ground states between each other are common (although not necessary). Having $\Psi$ and the resulting degeneracies for {\em all} the states is thus a much stronger condition, hence the name \cite{Alicea:2015}. 

A remarkable fact is that the strong zero mode and the degeneracies survive throughout the ordered phase $|\Gamma|<|J|$ of the Ising chain. This is quite simple to derive following the iterative procedure described in \cite{KitMajorana,Fendley:2016}, and we review this quickly here. We set $\Psi^{(0)}=\sigma^z_1$, since this commutes with $H_0$. However, it does not commute with the full Hamiltonian:
\begin{align} [H, \Psi^{(0)}] = 2i\Gamma\, \sigma^y_1\ .
\label{HPsi0}
\end{align}
We then need to add a term to the zero mode of order $\Gamma$ to cancel this; $\Psi^{(1)}=(\Gamma/J)\sigma_1^x\sigma^z_2$ does the trick. However, this now generates a term of order $\Gamma^2$:
\[ [H, \Psi^{(0)}+\Psi^{(1)}] = 2i(\Gamma^2/J)\, \sigma^x_1\sigma^y_2\ .\]
This in turn can be canceled by including a term $\Psi^{(2)}=(\Gamma/J)^2\sigma_1^x\sigma^x_2\sigma^z_3$, generating a new term of order $\Gamma^3$. 
Continuing in this fashion gives
\begin{align}
\nonumber
  \Psi  &=\mathcal{N}_{\rm Ising} \left[ \sigma^z_1 +  \frac{\Gamma}{J} \sigma^x_1 \sigma^z_2 +  \left(\frac{\Gamma}{J}\right)^2 \sigma^x_1 \sigma^x_2 \sigma^z_3 +\ldots \right]\\
 &=  \mathcal{N}_{\rm Ising} \sum_{j=0}^L \left(\frac{\Gamma}{J}\right)^{j}\sigma^z_j\prod_{k=1}^{j-1}\sigma^x_k\ .
 \label{eq:pureisingpsi} 
  \end{align}
Those familiar with the Jordan-Wigner transformation will recognize each term as a Majorana fermion. 

The operator $\Psi$  satisfies all three conditions for the strong zero mode throughout the ordered phase. Since a single $\sigma_j^z$ appears in each term, it anticommutes with ${\cal F}$.
Because each term in the expansion (\ref{eq:pureisingpsi}) anticommutes with each other one, setting the normalization to obey
\begin{align}
({\cal N}_{\rm Ising})^2= \frac{1-(\Gamma/J)^2}{1-(\Gamma/J)^{2L}} \approx 1-\left(\frac{\Gamma}{J}\right)^2\ .
\label{NIsing}
\end{align}
makes $\Psi^2=\mathds{1}$. The norm is non-vanishing as $L\to \infty$ when $|\Gamma|<|J|$. 
In this phase, $\Psi$ does indeed commute with the Hamiltonian up to an exponentially small correction:
\begin{equation}
{\mathcal E} \equiv [H, \Psi] = 2 \mathcal{N}_{\rm Ising} \Gamma  \left(\frac{\Gamma}{J}\right)^{L-1}{\cal F}\, \sigma^z_L.
\label{errorIsing}
\end{equation}
Moreover, in this phase, the norm of each term in the expansion (\ref{eq:pureisingpsi}) decreases quickly with $j$, justifying the name of edge strong zero mode. We of course can construct another strong edge zero mode on the other end, by starting with $\sigma^z_L$ instead of $\sigma^z_1$. The two edge zero modes anticommute with each other, indicating that they are indeed Majorana fermionic.

An intuitive way of thinking about the higher terms in the expansion of the strong zero mode is that they describe how information initially stored on the boundary ``leaks'' into the bulk. All the higher-order terms contain $\sigma^x_1$ and so flip the edge spin. Thus they partially, but not completely, decohere the edge spin. Indeed, their presences reduces $\mathcal{N}_{\rm Ising}$, and we will see in the next subsection \ref{sec:isingbm} how this reduces the asymptotic value of $A_s(\infty)$.

This computation of the edge strong zero mode is easy because the Ising chain can be mapped on to a free-fermion model. This result is however not a free-fermionic fluke. An analogous operator in the XYZ spin chain was found by this iterative method in \cite{Fendley:2016}. The XYZ chain has Hamiltonian
\begin{equation}
  \label{eq:XYZ}
 H_{\rm XYZ} = \sum_{j=1}^{L-1}\left[J_x \sigma^x_j \sigma^x_{j+1} + J_y  \sigma^y_j  \sigma^y_{j+1} +J_z \sigma^z_j  \sigma^z_{j+1}   \right]
\end{equation}
acting on the $L$-site chain with open boundary conditions. 
The spin-flip symmetry ${\cal F}$ commutes with $H_{\rm XYZ}$, so again the eigenstates can be grouped into two sectors. The extreme case $J_x=J_y=0$ reduces to the $\Gamma=0$ case of Ising, so this suggest iterating starting with $\sigma^z_1$. This yields
\begin{align}
\nonumber
\Psi_{\rm XYZ} &= {\cal N}_{\rm XYZ}\Big(\sigma^z_1  -\frac{J_y}{J_z}\sigma^x_1\sigma_2^x\sigma^z_3
-\frac{J_x}{J_z}\sigma^y_1\sigma_2^y\sigma^z_3\\
 &\quad+ \frac{J_xJ_y}{J_z^2}\big(\sigma^z_2-(\sigma^x_1\sigma_3^x+\sigma^y_1\sigma_3^y)\sigma^z_4\big) + \dots
\Big)
\label{PsiXYZ}
\end{align}
The explicit all-orders expression can be found in \cite{Fendley:2016}. Despite the many terms in this operator, $ \Psi_{\rm XYZ}^2=\mathds{1}$ for
\begin{equation}
\label{eq:XYZnorm}
\mathcal{N}_\text{XYZ} \approx \sqrt{\left(1-\frac{J_x^2}{J_z^2}\right)\left(1-\frac{J_y^2}{J_z^2}\right)}
\end{equation}
Thus $\Psi_{\rm XYZ}$ is normalizable for $|J_x|<|J_z|$ and $|J_y|<|J_z|$. 

The XYZ chain is interacting but integrable \cite{Baxbook}, possessing an extensive number of conserved quantities. This existence of the strong zero mode is clearly not independent of this fact, but the jury is still out on whether strong zero modes can be constructed for all integrable models. 

We stress that the pairing arising from a strong edge zero mode is a stronger statement than simply that the states are are degenerate or that they have finite overlap. For example, in a system with no hopping at all but non-zero on-site energies, such as in some effective Hamiltonians for many-body localised systems, a single spin operator would pair eigenstates in the sense of finite overlap. However, they would be separated by a (possibly large) on-site energy. Conversely, even if the pairing due to an edge mode starting at $\sigma^z_1$ were to break down, there is no guarantee that the system would no longer have degenerate paired energy levels due to some other, not edge-localised, zero mode.

\subsection{Edge spin coherence}
\label{sec:isingbm}

An infinitely long coherence time for the edge spin results when energy eigenstates in different sectors pair up as described in section \ref{sec:basic}. The strong zero mode guarantees both conditions for the pairing. Moreover, knowing the strong zero mode explicitly allows us to compute the leading contribution to the asymptotic values $A_s(t\to\infty)$. We find that not only is this value non-vanishing, but independent of the state $s$, even if highly excited.

Precisely, all eigenstates $|s_\pm\rangle$ obeying ${\cal F}|s_\pm\rangle=\pm|s_\pm\rangle$ pair up via
\begin{align}\label{eq:pairing}
  \ket{s_\pm} &\approx\Psi \ket{s_\mp} 
\end{align}
where as always $\approx$ means up to terms exponentially small in $L$. The matrix element $\Braket{s_+|\sigma_1^z|s_-}$ is non-vanishing as a consequence.
Namely,
\begin{align}
\nonumber \Braket{s_\pm|\sigma_1^z|s_\mp}  & =  \frac{1}{2} \left(  \Braket{s_\pm|\sigma_1^z|s_\mp}+ \Braket{s_\mp|\sigma_1^z| s_\pm} \right) \\
 &
  \approx  \frac{1}{2} \Braket{s| \{\Psi, \sigma_1^z\}| s}\ ,
  \label{matrixanti}
\end{align}
where we exploited the fact that $\sigma^z_1$ is hermitian. However, the leading term in $\Psi$ is simply ${\cal N}\sigma^z_1$, where ${\cal N}$ is the normalization making $\Psi^2=\mathds{1}$. 
Thus 
\begin{align}
\Braket{s^\prime|\sigma_1^z|s} = \mathcal{N} + \hbox{ corrections }.
\label{ssN}
\end{align}
The order of the corrections depends on the model. 
For Ising, the corrections are exponentially small (of order $1/J^L$), because all the terms in $\Psi$ given by~\eqref{eq:pureisingpsi} anticommute with $\sigma^z_1$ except $\sigma^z_1$ itself. For XYZ, the corrections occur at lowest at order $1/J_z^2$. This is because the next term in the explicit expansion (\ref{PsiXYZ}) of $\Psi$ not anticommuting with $\sigma_1^z$ is proportional to $\sigma^z_2$, occuring at order $1/J_z^2$.

We have thus shown when there is a strong zero mode, the magnitude of the boundary magnetization is not only non-vanishing, but independent of $s$ at leading order. Since $E_{s_+}\approx E_{s_-}$, (\ref{ssN}), this also guarantees that the coherence time is infinite when $L\to\infty$.
It is of course possible that other states of close energy have finite overlap and contribute to (\ref{Asum}), but these will only increase it further, since all terms in the sum only involve the magnitude of the overlap. It is also possible that there could be finite overlap between states of different energy, thus contributing oscillating terms not vanishing as $L\to\infty$ and $t\to\infty$. If we ignore such additional contributions, we then find for Ising
\begin{align}
A_s(\infty) \equiv\lim_{t\to\infty}\lim_{L\to\infty} A_s(t) \approx {\cal N}_{\rm Ising}^2 \approx 1-\frac{\Gamma^2}{J^2}
\label{Isinginfty}
\end{align}
For the ground states, this was derived long ago \cite{Pfeuty}. However, the fact that it holds for all states with an $s$-independent value was not previously noted, as far as we are aware.

\begin{figure} [htbp]
\centering
 \includegraphics[width=\linewidth]{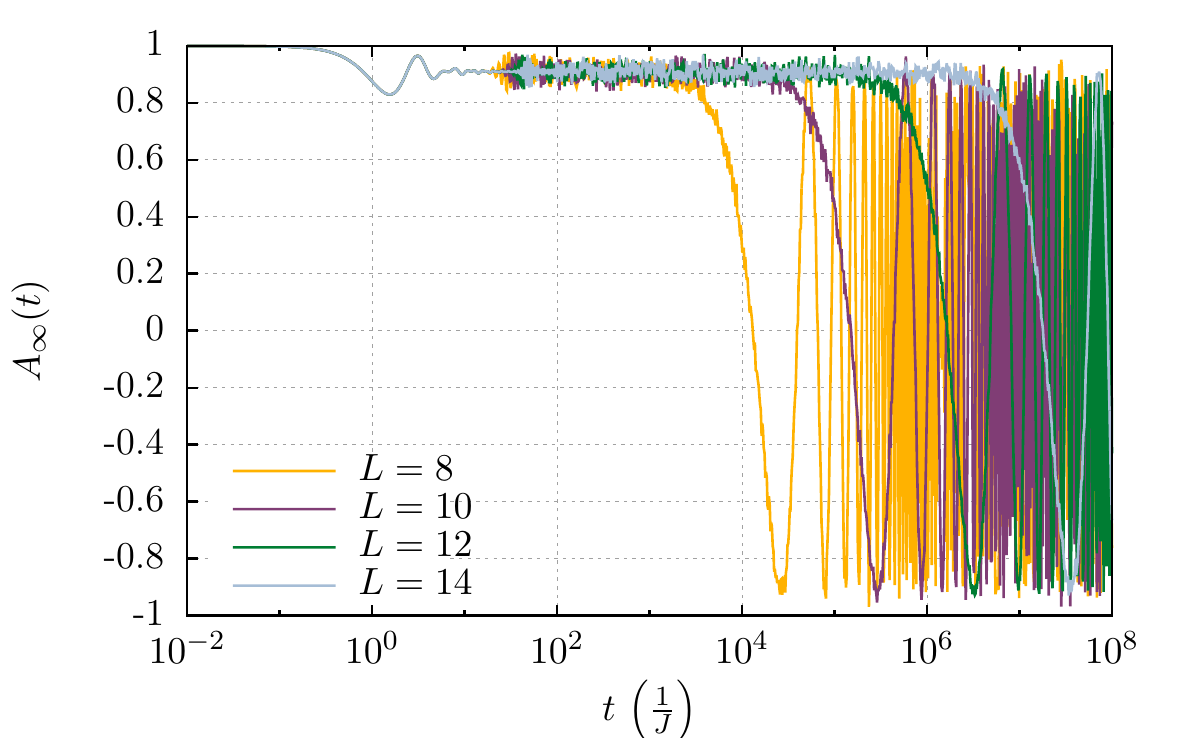}
\caption{$A_\infty(t)$ from exact diagonalization on a log axis for time for $\Gamma = 0.3$. Revivals appear at a time scale set by the finite size.}
\label{fig:Ising_log_decay}
\end{figure}

Since (\ref{Isinginfty}) is independent of $s$, it holds true with any initial state, not just eigenstates of $H_{\rm Ising}$. Even the  ``infinite-temperature'' autocorrelator  is non-vanishing. With a slight abuse of notation, we dub this autocorrelator as
\begin{align}
A_\infty(t)\equiv \frac{1}{2^L}\sum_s A_s(t)\ .
\end{align}
Using exact diagonalization, $A_\infty(t)$ is plotted for $\Gamma=0.3$ for very long times in figure 
\ref{fig:Ising_log_decay}; note that the time axis is logarithmic.
We see that initial oscillating pieces die off very quickly in time to a long-lived plateau at the value ${\cal N}_{\rm Ising}^2 = (1-(.3)^2)=.91$. The coherence time indeed grows exponentially with increasing $L$, since the error term ${\cal E}$ has norm decaying exponentially. The plateau falls off at a time roughly $1/|{\cal E}|$, where the decay time arises from the ``error'' in (\ref{errorIsing}). This gives the finite-size dependence of the coherence to be
\begin{align}
T_{\rm Ising} \sim \frac{1}{\sqrt{J^2-\Gamma^2}}\left(\frac{J}{\Gamma}\right)^L\ .
\label{TIsing}
\end{align}
This indeed goes to infinity as $L\to\infty$, but even for rather modest system sizes, the finite-size effects do not appear until very long times. 

An amusing phenomena apparent at finite sizes and long enough times is that the autocorrelator ``revives'', returning to its initial value. That this is visible in such a plot is a consequence of the free-fermion nature of Ising: the energy differences in all Ising pairs are identical, and so sum up coherently at much shorter times than they do in an interacting system such as XYZ, illustrated below. (This fact is also the reason for the persistence of the wiggles on each plateau.)
Since Ising can be solved, the time for the revivals can be computed; this is easily done using the setup of section 2.2 of \cite{freepara}.  One finds that at finite size each pair $s_\pm$ are split in energy by the identical amount $\epsilon=\Gamma(\Gamma/J)^{L-1}$ at leading order in $\Gamma/J$. This implies a revival time of order $1/\epsilon$, where the oscillating pieces are back in sync again. Thus in Ising, there really is no decay of the plateau: the revival time is of the same order as the coherence time, as indeed apparent from figure \ref{fig:Ising_log_decay}.

\begin{figure} [htbp]
\centering
 \includegraphics[width=\linewidth]{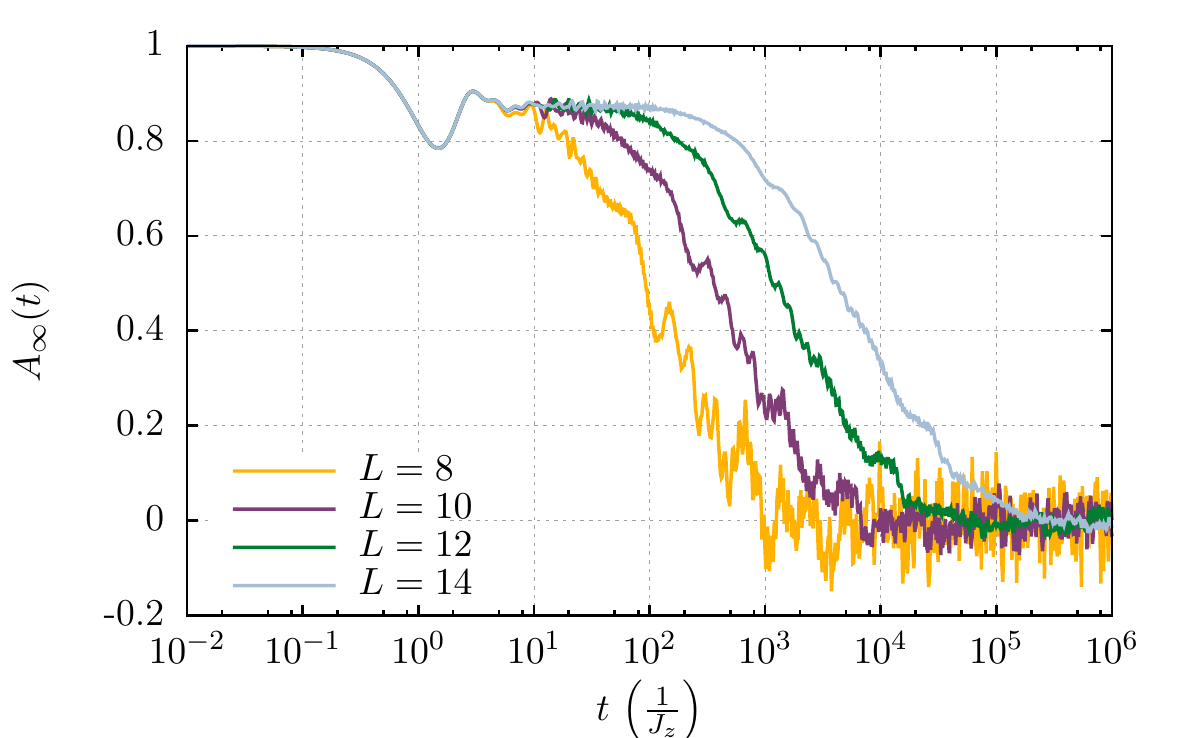}
\caption{$A_\infty(t)$ for XYZ from exact diagonalization on a log axis for time for
$L = 14$, $J_x = 0.2$, $J_y =0.3$, $J_z = 1$.}
\label{fig:XYZ_log_decay}
\end{figure}

\begin{figure} [htbp]
\centering
 \includegraphics[width=\linewidth]{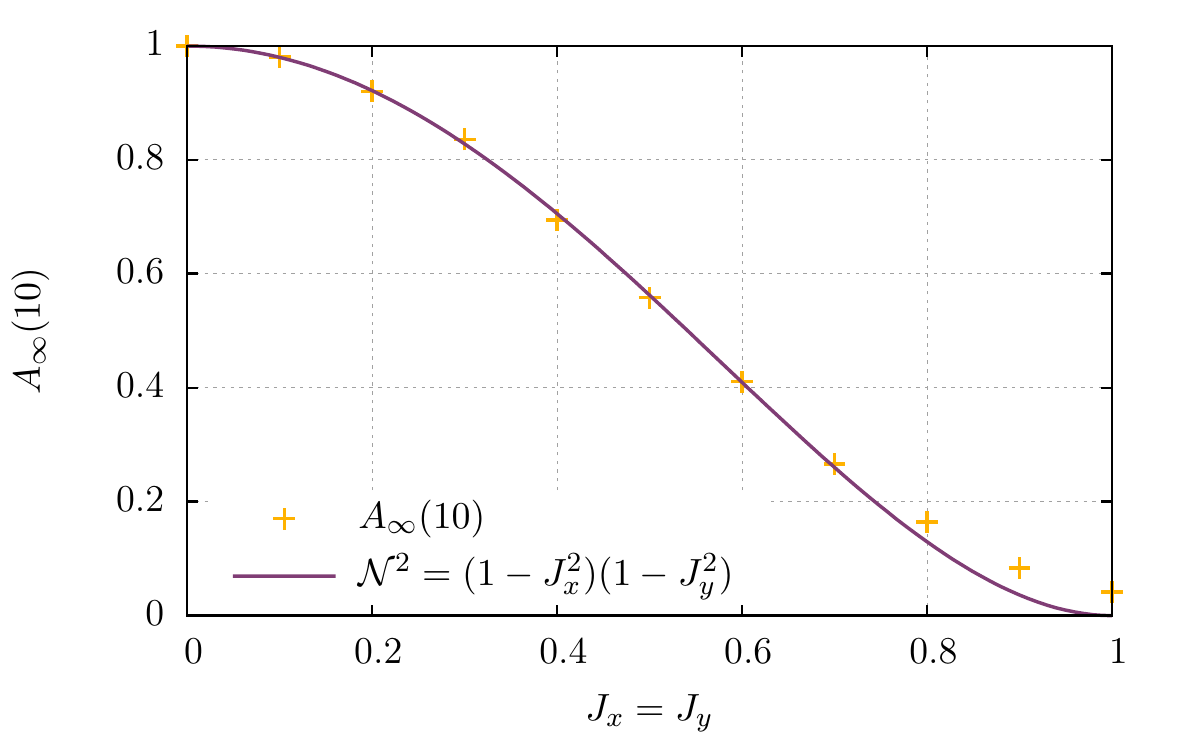}
\caption{$A_\infty(10)$ for XYZ from exact diagonalization as a function of $J_x=J_y$ for $J_z = 1$ and $L = 14$, compared with the analytic estimate from (\ref{AinfXYZ}).}
\label{fig:XYZ_plateau}
\end{figure}

For the XYZ chain, we plot $A_\infty(t)$ from exact diagonalization in figure \ref{fig:XYZ_log_decay}, and indeed see the plateaus. 
Here the leading contribution to autocorrelator at large times is
\begin{align}
A_s(\infty) \approx {\cal N}_{\rm XYZ}^2 \approx \left(1-\frac{J_x^2}{J_z^2}\right)\left(1-\frac{J_y^2}{J_z^2}\right)
\ .
\label{AinfXYZ}
\end{align}
We defer a careful discussion of the corrections to section \ref{sec:error}; for $A_{\infty}$ they turn out to be exponentially small as in Ising.
The plateau values obtained from the numerics are very close to the those in (\ref{AinfXYZ}), as is clear from figure \ref{fig:XYZ_plateau}. This gives 
a strong sign that there is no pairing occurring other than that from the edge mode. Even though the XYZ chain is integrable, it is not free-fermionic. Thus the energy splittings are not related, and revivals do not appear in $A_\infty$. 

The fact that the edge coherence time is infinite for {\em all} states is by no means an obvious consequence of integrability. In an integrable model, there are an extensive set of conserved quantities. Whereas boundary conditions could wreck these, this does not happen for open boundary conditions (often known in the literature as ``free'') in Ising or XYZ. Certainly the presence of these conserved quantities is related to the presence of the strong zero mode and the infinite coherence time. However, they do not guarantee that any spin, edge or otherwise, has such coherence. Indeed, if one attempted to construct a strong zero mode starting with the operator $\sigma^z_k$ with $k$ in the bulk of the sample, the procedure quickly breaks down \cite{Ros:2014} because of the resonances described in section \ref{sec:resonances}. Thus in spite of the integrability,  in a highly excited eigenstate the spins in the bulk of the system very quickly appear as if they thermalize, losing all coherence. 

Note also that the strong edge zero mode only approximately commutes with the Hamiltonian in finite size, as opposed to the usual conserved quantities arising from integrability. In fact, it is not at all obvious how it appears inside the usual frameworks of integrability, such as the Bethe ansatz. Possibly it is related to the appearance of boundary bound states, as discussed in \cite{Skorik:1995,Kapustin:1995}. It would be very interesting both for these studies and for those of integrable models to understand better how all of this structure fits together.

\section{Long coherence time from an almost strong zero mode}
\label{sec:aszm}

Here we show how long but finite coherence times arise. We study the Ising chain (\ref{HIsing}) modified by including two additional types of terms, a nearest-neighbor $\sigma^x$ coupling, and a next-nearest-neighbor $\sigma^z$ coupling. The Hamiltonian is
\begin{align}H=H_{\rm Ising}-\Gamma_2\sum_{j=1}^{L-1}\sigma^x_j\sigma^x_{j+1}- J_2\sum_{j=1}^{L-2}\sigma^z_j\sigma^z_{j+2}\ .
\label{Hpert}
\end{align}
In the fermionic version, the new terms are the simplest four-fermion interactions, but as opposed to the XYZ chain, either interaction breaks the integrability.  Both perturbing operators have dimension 4 at the Ising critical point,  and so are irrelevant in the renormalization-group sense. We thus expect that for small enough $J_2$ and $\Gamma_2$, the coherence time should remain infinite for all eigenstates $s$ whose energy density is the same as that of the ground state. This is not surprising, since if all the couplings other than $J$ are small, the model is ordered.

\begin{figure} [htbp]
\centering
 \includegraphics[width=\linewidth]{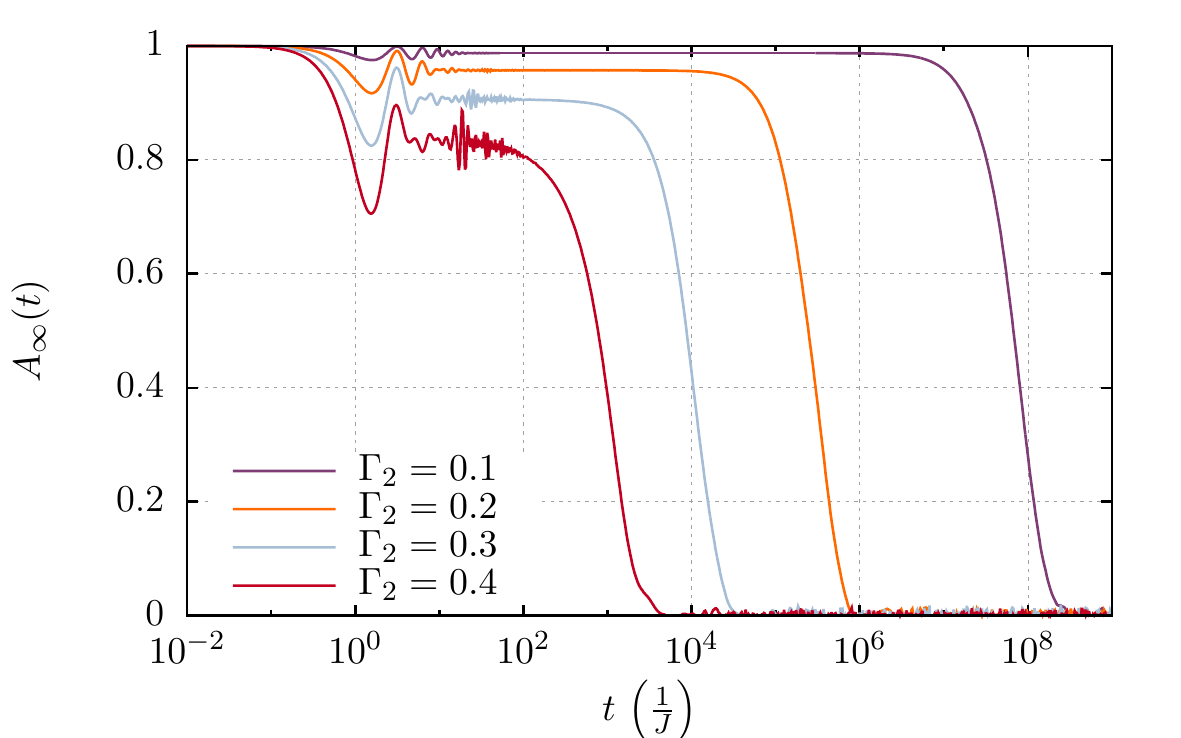}
\caption{$A_\infty(t)$ for perturbed Ising from exact diagonalization on a log axis for time for
$L = 14$, $\Gamma = 0.05$, $J_2=0$.}
\label{fig:vdecay}
\end{figure}

The consequences for the excited states, however, are much more surprising.  We plot $A_\infty(t)$ with a log time axis in figure~\ref{fig:vdecay} for several small values of $\Gamma_2$, with $J=1$ and $\Gamma= 0.05$. 
While $A_s(t)$ does eventually approach zero, it takes a very long time to do so. For small $\Gamma_2$, the edge spin coherence time can reach millions of time steps (in units of $1/J$), even for the system sizes accessible by exact diagonalization.

\begin{figure} [htbp]
\centering
 \includegraphics[width=\linewidth]{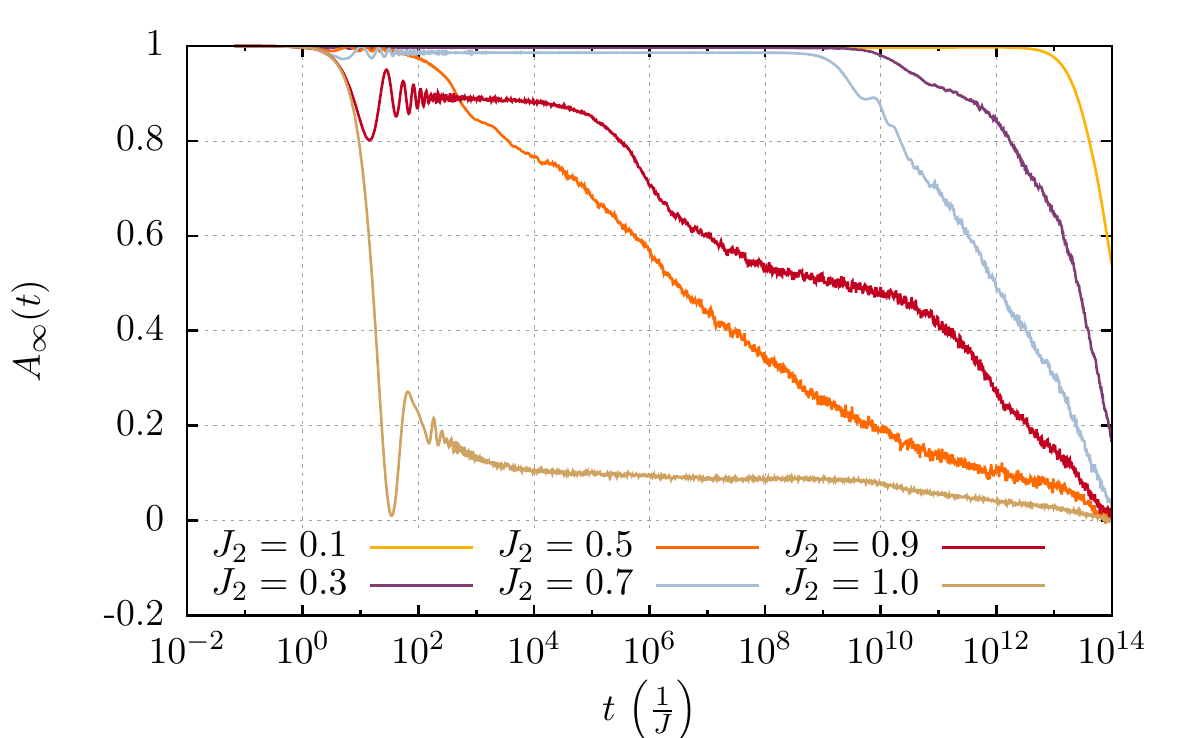}
\caption{$A_\infty(t)$ for perturbed Ising from exact diagonalization on a log axis for time for
$L = 14$, $\Gamma = 0.05$, $\Gamma_2=0$. Notice the non-monotonicity in $J_2$.}
\label{fig:j2decay}
\end{figure}

An even richer story arises for $J_2\ne 0$. At small $J_2$, the behavior is similar to that for small $\Gamma_2$. However, for larger values, the behavior is completely different, as is clear from figure~\ref{fig:j2decay}. The edge coherence time does {\em not} gradually decrease as $J_2$ increases.  For example, at $J_2=.5J$, the plateau is short-lived, whereas at $J_2=.7J$ the decay time is similar to that at $J_2=.3J$, in the billions of time steps. At $J_2=J$, the decay is quick.  Furthermore, the behaviour of $A_\infty(t)$ is qualitatively the same for {\em negative} $J_2$, despite there being frustration in the low-lying states. Clearly there is more to the story than just integrability breaking.

\subsection{Iterating and Normalizing}
\label{sec:itnorm]}

Given the results for Ising and XYZ, the obvious next step is to try using the iterative procedure to find a strong zero mode.  Consider first the case with $J_2 =0$. The $\Gamma_2$ coupling is a disordering term like $\Gamma$, and does not commute with $J$. In fact, if $\Gamma=0$ as well, then (\ref{Hpert}) reduces to the XYZ chain with $J_y=0$ and $\Gamma_2=J_x$. It is thus natural to treat $\Gamma_2$ as a further perturbing term on the same order as $\Gamma$, and indeed to first order in perturbation theory the pairing between sectors persists \cite{Kells}. In both Ising and XYZ cases, the strong zero mode starts with $\Psi^{(0)}=\sigma^z_1$. Starting with that here means
the first correction must be the sum of those in Ising and XYZ:
\[ \Psi^{(1)} = \frac{\Gamma}{J} \sigma_1^x\sigma^z_2-\frac{\Gamma_2}{J}\sigma^y_1\sigma^y_2\sigma^z_3\ .\]
However, now $[H,\Psi^{(0)}+\Psi^{(1)}]$ is considerably more complicated than in the Ising or XYZ cases, containing five operators proportional to $\Gamma\Gamma_2$. It however, is still possible to find a $\Psi^{(2)}$ that cancels all these terms. It is
\begin{align}
\nonumber
\Psi^{(2)} =&- \frac{\Gamma \Gamma_2}{J^2} \left( \sigma^x_{1} \sigma^y_{2} \sigma^y_{3} \sigma^z_{4} +\sigma^x_{1} \sigma^z_{3} +\sigma^x_{2} \sigma^z_{3} + \sigma^y_{1} \sigma^y_{2}\sigma^x_{3} \sigma^z_{4}\right)\\
&\quad+\frac{\Gamma^{2}}{J^2} \sigma^x_{1} \sigma^x_{2} \sigma^z_{3} 
+\frac{\Gamma_2^{2}}{J^2} \sigma^y_{1} \sigma^y_{2} \sigma^y_{3} \sigma^y_{4} \sigma^z_{5} \ .
\label{Psitwo}
\end{align}
Then $[H,\Psi^{(0)}+\Psi^{(1)}+\Psi^{(2)}]$ is of third order in $\Gamma,\Gamma_2$.

Despite how complicated this is, the procedure still works. At each step, one must solve the equation
\begin{align}
[H,\Psi^{(0)}+\Psi^{(1)}+\dots+\Psi^{(n-1)}]=-J[V,\Psi^{(n)}]\ ,
\label{szmrecursion}
\end{align}
where $V=\sum_j\sigma^z_j\sigma^z_{j+1}$.
The non-trivial step in finding $\Psi^{(n)}$ is therefore the inversion of the operation $[V, \cdot]$. There is no guarantee that it is invertible: the null space, the set of operators commuting with $V$, is exponentially large. For example, it includes any function of the $\sigma^z_j$. Moreover, even if a $\Psi^{(n)}$ can be found, it is not unique: any operator in this null space may be added onto $\Psi^{(n)}$ without changing its commutator with $V$. This turns out to be quite a feature, because such an addition will not commute with the full $H$. Thus it does change the equation that determines $\Psi^{(n+1)}$, allowing one to search for a particular choice that makes $[V,\cdot]$ invertible here.  For example, for the XYZ case, this is responsible for the $\sigma^z_2$ term appearing in (\ref{PsiXYZ}).  We find that in perturbed Ising, like in XYZ,  there is a unique choice of such a term at order $n$ to make the equation for $\Psi^{(n+1)}$ solvable. 

With the help of a Python program to do the algebra, we have implemented this procedure successfully to 7\th order for $J_2=0$, 11\th order when $\Gamma_2=0$, and 6\th order when both $\Gamma_2$ and $J_2$ are included. The fact that this iteration is possible is highly non-trivial; as mentioned above, if one starts iterating instead with $\sigma^z_j$ with $j\ne 1$ or $L$, the procedure breaks down instantly. Although starting at the edge works, the number of terms explodes dramatically compared to the integrable cases. At 8th order, there are 68,368 terms in the expansion, as compared to 9 terms at the same order for Ising. 

\begin{figure} [htbp]
\centering
 \includegraphics[width=\linewidth]{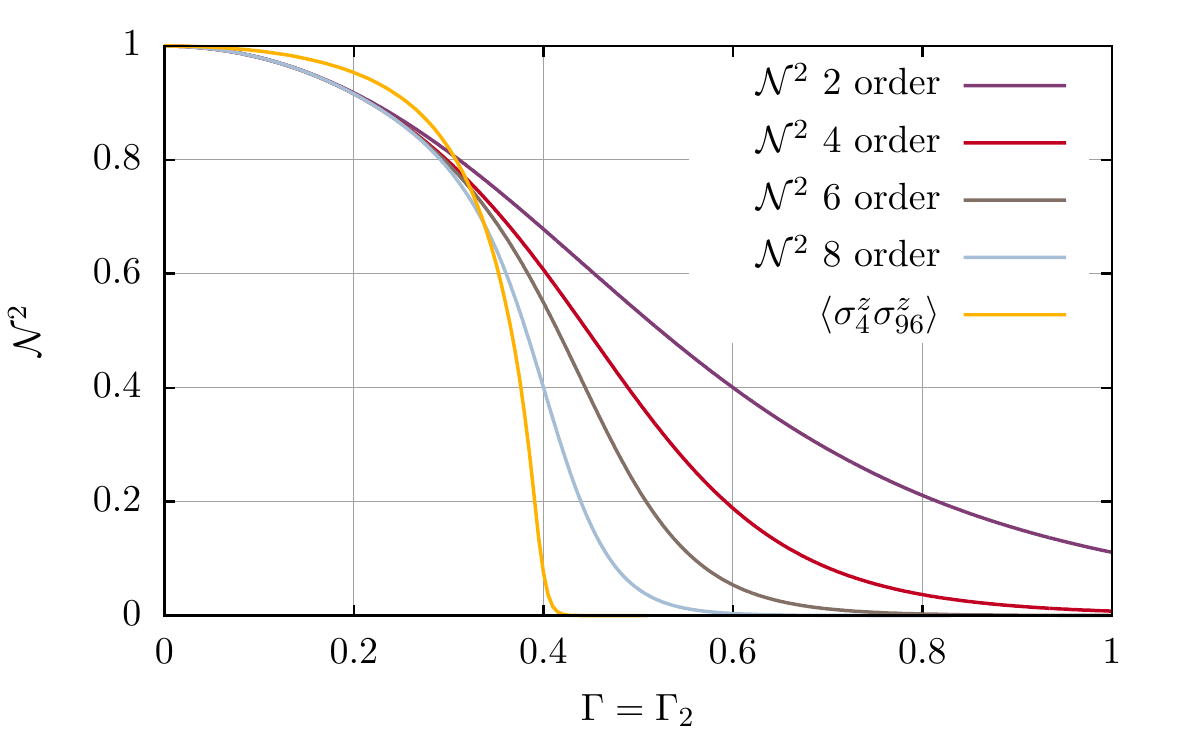}
\caption{The normalisation $\mathcal{N}$ for $\Gamma=\Gamma_2$ and $J_2=0$ plotted at different truncation orders, compared with the two-point correlator in the ground state from DMRG with $L=100$.}
\label{fig:norm_vs_trans}
\end{figure}

Another criterion for a strong zero mode is that $\Psi^2=\mathds{1}$. Again using the computer to do the algebra, we find that to the order we know $\Psi$, the square $\Psi^2$ is indeed proportional to the identity operator. This quite remarkable cancellation allows us to to define the normalization as the coefficient ${\cal N}$ that makes
${\cal N}(\Psi^{(0)} + \Psi^{(1)}+\dots)$ square to $\mathds{1}$. 
To fourth order 
\begin{align*}
  \mathcal{N}^{-2} &=1 + \Gamma^{2} + \Gamma_2^{2}  + \Gamma^{4} + 3\Gamma^{2} J_{2}^{2} - 8\Gamma_{1} \Gamma^{2} J_{2} \\ & \qquad+  4 \Gamma_2^{2} \Gamma^{2}+ 6 \Gamma_{2}^{2} J_{2}^{2} +\Gamma_2^4 
 + \dots\ ,
\end{align*}
where $J_2=1$.
The coefficients increase further at higher orders. The normalization as a function of $\Gamma=\Gamma_2$ and $J_2=0$ for various truncations is plotted in figure \ref{fig:norm_vs_trans}. We also plot the two-point correlator in the ground state, which shows the usual quantum phase transition occurs at $\Gamma=\Gamma_2\sim 0.4$. We see that it is plausible that the norm remains non-zero even as $L\to\infty$ throughout the ordered phase, although we stress that this is not necessary for any of our arguments below to hold. 

\subsection{The error term}
\label{sec:error}

The preceding gives good evidence that despite the enormous number of terms, the iterative procedure works for the perturbed Ising model. At every order, one presumably can find another term that cancels the error when commuted with $V$. This very possibly yields a normalizable operator that squares to the identity. However, it does not guarantee that the norm of the error term vanishes exponentially with system size, as is necessary to get the pairing. It is straightforward to see that this does {\em not} happen here. In fact, it does not go to zero at all, but instead there exists some order $n_*$ (depending on the couplings) where
including $\Psi^{(n_*+1)}$ and higher terms in the expansion increases the error. 

We define the error at each order as
\begin{align}
{\cal E}_n= [H,\Psi^{(0)}+\Psi^{(1)}+\dots+\Psi^{(n-1)}]\ .
\label{errorn}
\end{align}
This of course is the error used to determine the strong zero mode at next order in the recursion relation (\ref{szmrecursion}). We then define $n_*$ as the lowest value such that
\begin{align}
|{\cal E}_{n_*+1}| >|{\cal E}_{n_*}|\ ,
\label{errornstar}
\end{align}
where we use the trace norm $|{\cal E}_n|^2\equiv\Tr({\cal E}_{n}^\dagger{\cal E}_{n})$.
Under this definition,  $n_* \to \infty$ for Ising.

\begin{figure} [htbp]
\centering
 \includegraphics[width=\linewidth]{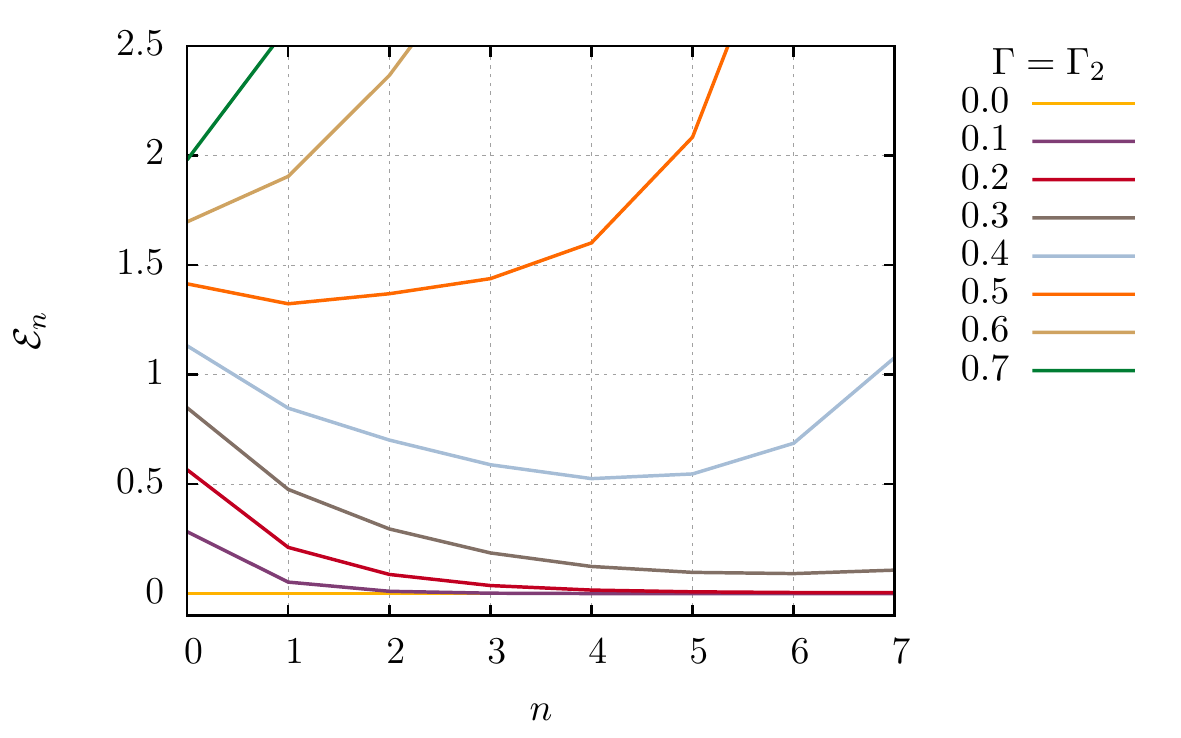}
\caption{$|{\cal E}_{n}|$ as a function of truncation order $n$.}
\label{fig:gamma2_error_f_n}
\end{figure}

Using the explict expressions for $\Psi^{(n)}$ allows us to compute the error ${\cal E}_n$ to the same order. We find that when $J_2=0$, the maximum errors occur roughly at $\Gamma_2=\Gamma$, not surprisingly as the strong zero mode is exact for $\Gamma\Gamma_2=0$.   We thus plot ${\cal E}_n$ for various $\Gamma=\Gamma_2$ in figure~\ref{fig:gamma2_error_f_n}; the lines are there to make the trends clear. 
The truncation point $n_*$ for given couplings is simply the minimum of the corresponding curve as a function of $n$. For example, when $\Gamma=\Gamma_2\approx 0.25$ the expansion must be truncated at $n_*=7$. Plotting the same graph for $J_2 = \Gamma$ and $\Gamma_2=0$ gives qualitatively similar results.

\subsection{The almost strong zero mode}
\label{sec:almost}

There is no strong zero mode in the perturbed Ising model. This is in accord with the numerics: a strong zero mode would have required an infinite coherence time as $L\to\infty$. However, not only does the iterative procedure work, but it yields an operator that such that when squared, thousands of complicated terms cancel, giving the identity to the order truncated. This striking behavior hardly seems meaningless. We thus define an  {\em almost strong zero mode} by truncating the expansion at $n_*$:
\begin{align}
\Psi_*={\cal N}_*\sum_{n=0}^{n_*} \Psi^{(n)}\ ,
\label{Psistar}
\end{align}
where ${\cal N}_*$ is determined by requiring $\Psi_*^2=\mathds{1}$. 
By generalizing the arguments in section \ref{sec:isingbm}, we show here that the almost strong zero mode implies the long but finite coherence time seen in the numerics. Moreover, $({\cal N}_*)^2$ is the leading contribution to the height of the plateau in $A_\infty(t)$. 

Since the error term does not vanish, the eigenstate pairs with finite overlap from $\sigma^z_1$ no longer survive the $L\to\infty$ limit.  However, the almost strong zero mode still results in relations between different states (but not both eigenstates). 
We define a {\em partner} $|\psi_s\rangle$ for each eigenstate $|s\rangle$ by
\begin{align}\ket{\psi_s}\equiv \Psi_*|s\rangle\ .\end{align}
As in (\ref{matrixanti}), the partner has non-vanishing overlap with $|s\rangle$: 
\begin{align}
\nonumber
\Braket{\psi_s|\sigma_1^z|s} &=  \frac{1}{2} \Braket{s| \{\Psi_*, \sigma_1^z\}| s}\\
& = \mathcal{N}_* + \hbox{ corrections }.
\label{partneroverlap}
\end{align}
The partners are the paired states when $n_*\to\infty$, but when $n_*$ is finite, they are not: 
the partner $\psi_s$ is {\em not} an exact eigenstate of $H$ because of the non-vanishing error term. The important fact though is that the corrections are at order $n_*$:
\begin{align*}
H\ket{\psi_s}&=H\Psi_*\ket{s}=E_s\ket{\psi_s}+{\cal E}_{n_*}\ket{s}\sim E_s\ket{\psi_s}\ ,
\end{align*}
where $\sim$ means up to terms of order $1/J^{n_*+1}$. Having $(\Psi_*)^2\sim \mathds{1}$ means that $\Braket{\psi_{r}|\psi_s}\sim \delta_{rs}$ 
and the $\ket{\psi_s}$ form a complete linearly independent set of states.
Inserting this set into the expansion for $A_s(t)$ analogously to (\ref{Asum}) yields
\begin{align}
\nonumber
{A}_{s}(t)  &\sim \sum_r \Braket{s|\sigma^z_1| \psi_r}\Braket{\psi_r|e^{-i H t}\sigma^z_1e^{i H t}|s} \\
&\sim  \sum_r |\Braket{s |\sigma^z_1| \psi_r}|^2 e^{i\left(E_{s}-E_{\psi_r}\right)t}\ .
\label{Aspartner}
\end{align}

Neglecting the finite error terms means we have neglected energy differences of order $|{\cal E}_{n_*}|$. The consequence is that using (\ref{partneroverlap}) in (\ref{Aspartner}) implies a long but not infinite coherence time. For times long enough for the oscillating pieces to cancel, but shorter than $1/|{\cal E}_{n_*}|$, we expect a plateau with
\begin{align}
A_s\big|_{\rm plateau} \sim ({\cal N}_*)^2 + \hbox{ corrections}\ .
\end{align}
It follows from (\ref{Psitwo}) that here the corrections are of order $\Gamma\Gamma_2/J^2$. 
For $A_\infty(t)$, they are smaller. Using the Cauchy-Schwarz inequality gives
\[\sum_s|\Braket{s| \{\Psi_*, \sigma_1^z\}| s}|^2 \geq \frac{1}{2^L}\Tr(\{\Psi_*,\sigma^z_1\})^2\ .\]  This trace necessarily vanishes for all terms other than diagonal ones, because they are comprised of tensor products of Pauli matrices. The diagonal ones are positive, so the corrections to $A_\infty(t)$ are of order $n_*+1$:
\begin{align}
A_\infty\big|_{\rm plateau} \gtrsim ({\cal N}_*)^2 \ .
\end{align}

\begin{figure} [htbp]
\centering
 \includegraphics[width=\linewidth]{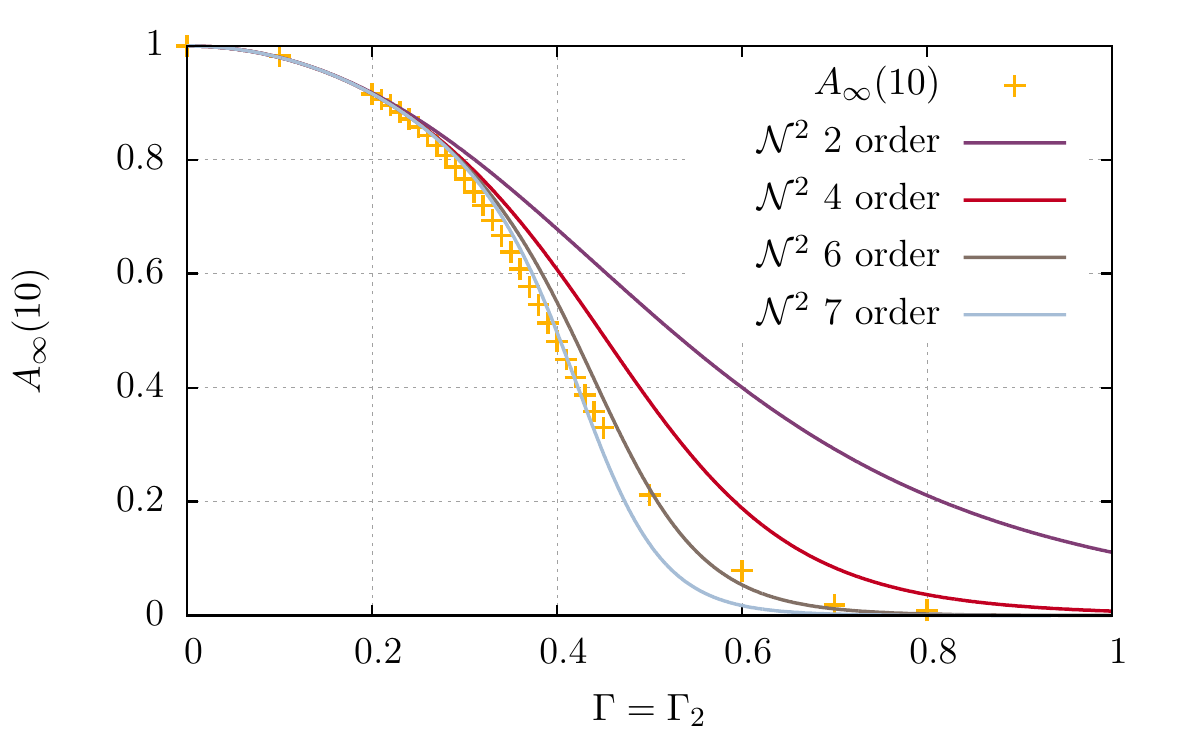}
\caption{$A_\infty (10)$, the plateau height, for $L=14$ and $J_2=0$ from exact diagonalization, compared with the estimate.}
\label{fig:gamma2_plat}
\end{figure}

\begin{figure} [htbp]
\centering
 \includegraphics[width=\linewidth]{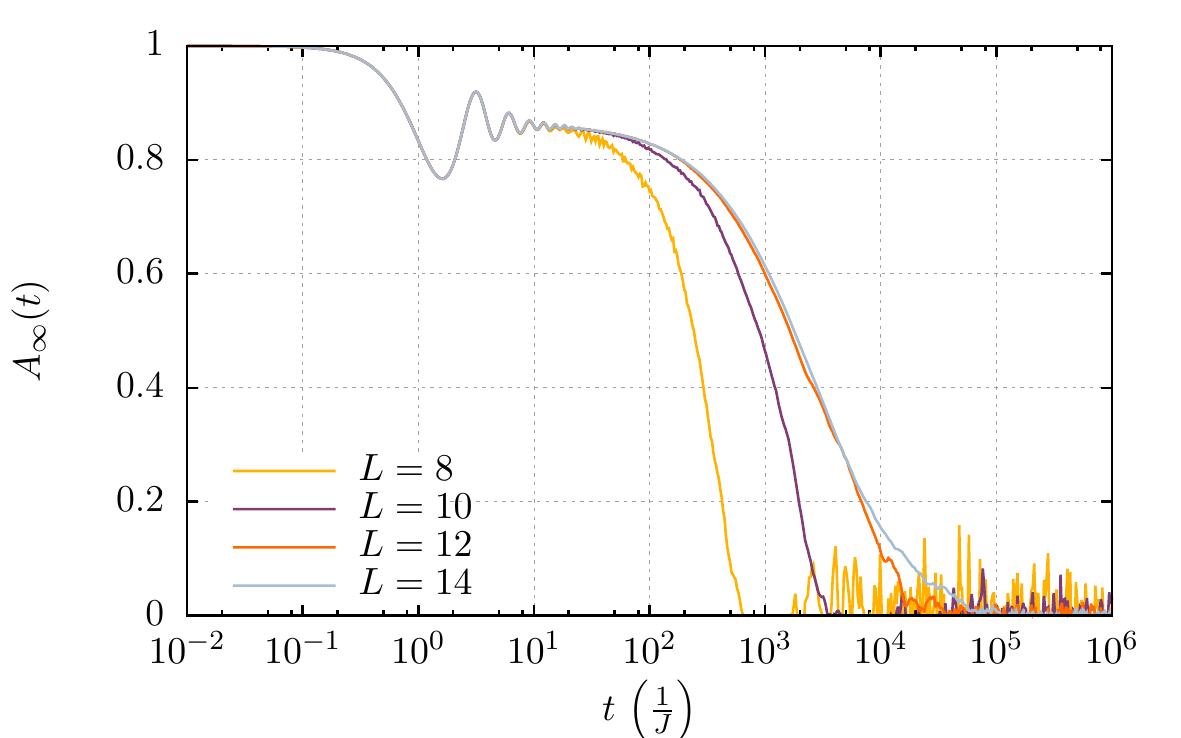}
\caption{$A_\infty (t)$ for Ising with $\Gamma = \Gamma_2 = 0.25$ from exact diagonalization. Notice the curve saturates with $L$.}
\label{fig:gamma2_sat}
\end{figure}

To check these assertions,  we compute $A_\infty(t)$ using exact diagonalization at $L=14$. As in the XYZ case in figure \ref{fig:XYZ_plateau}, we compare the plateau values with ${\cal N}_*^2$ for various $n_*$ in figure \ref{fig:gamma2_plat}, and find that the estimate is fairly accurate.
We find that typically the plateau is {\em longer} lived than $1/|{\cal E}_{n_*}|$. For example, for $\Gamma=\Gamma_2=0.25$, we see from figure 
\ref{fig:gamma2_error_f_n} that $|{\cal E}_{n_*}|\sim 0.025 J$, while it is clear from figure \ref{fig:gamma2_sat} that there is appreciable coherence at times longer than $1000/J$ steps. 

Another compelling check that the non-vanishing error term results in the finite decay time comes from showing $A_\infty(t)$ is independent of $L$ at large enough sizes. This requires care to see using exact diagonalization; because the support of $\Psi_*$ involves $2n_*+1$ spins, for most couplings finite-size effects cause the decay before the finite $n_*$ does.  Since $n_*=7$ for $\Gamma=\Gamma_2=0.25$, we plot $A_\infty(t)$ here in figure \ref{fig:gamma2_sat} for various system sizes. Happily we do see saturation: the curves for $L=12$ and $L=14$ are virtually identical. This is a good sign that the physics we are describing is not simply a small-sizes phenomenon.



\begin{figure} [htbp]
\centering
 \includegraphics[width=.95\linewidth]{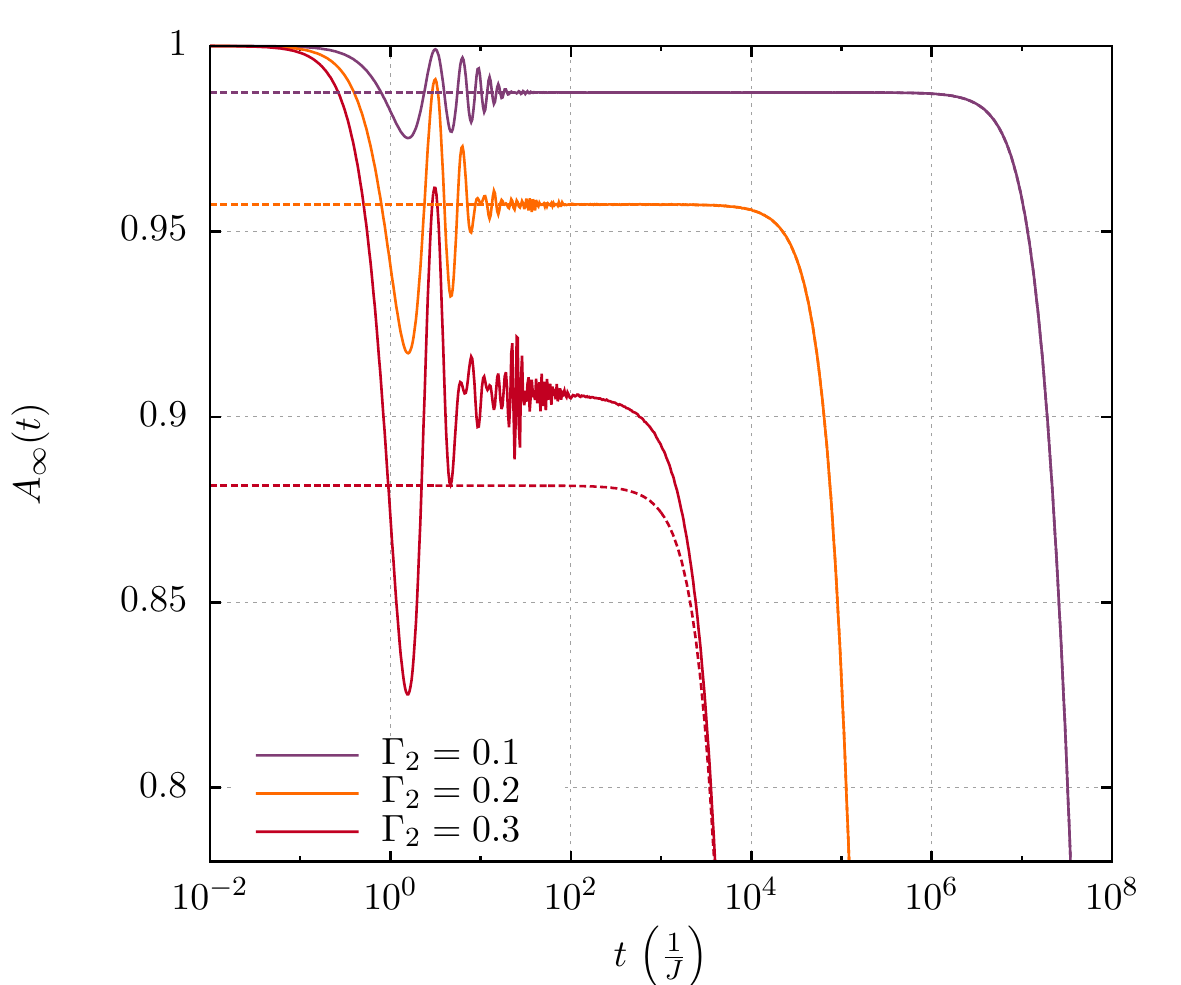}
\caption{A comparison of $A_\infty(t)$ calculated using all contributions (solid line) and only those from paired eigenstates (dashed line) for $L=14$, $\Gamma=0.05$.}
\label{fig:paireddecay}
\end{figure}

A long (albeit finite) edge coherence time thus is possible even without the exact pairing arising from the exact strong zero mode.  To check this scenario further, we numerically found all the energy eigenstates $s_\pm$. For each $s_+$, we then find the state $s_-$ with the maximum magnitude of the matrix element $\Braket{s_\mp|\sigma^z_1|s_\pm}$, and average this maximum over all states $s_+$. This clearly is heading to zero as $L$ increases, as the lack of a strong zero mode would imply. However, the long coherence time still survives. In figure~\ref{fig:paireddecay}, we plot both $A_\infty(t)$ and the contributions to it coming solely from the pairs of maximum overlap in the sum in (\ref{Asum}). 
Obviously, these two differ at very short times where the oscillating terms do not all cancel, but for small enough $\Gamma_2$, the two curves agree almost exactly at larger times. This indicates that at small system sizes and couplings, the dominant contribution to $A_\infty(t)$ indeed comes solely from the paired states. However it is also clear that for larger $\Gamma_2$ there are long-lived contributions which do not stem from the paired states. These presumably are the contributions coming from the partners as described above.

\section{Resonances and Poles}
\label{sec:resonances}

The two numerical checks presented in figures \ref{fig:gamma2_plat} and \ref{fig:gamma2_sat} provide convincing evidence that the almost strong zero mode is the physics underlying the long edge spin coherence time when $\Gamma_2$ is small. All the issues discussed in the previous section are also relevant to the $J_2\ne 0$ case as well. However, we find some additional interesting behavior here as well, such as that in figure \ref{fig:j2decay}, where the dependence of the decay time on $J_2/J$ clearly is not monotonic. 

We show here that this behavior is a consequence of the phenomenon described in \cite{Jermyn:2014,Ros:2014}: at certain values of $J_2/J$, it becomes ``easy'' to flip the boundary spin, and so the boundary coherence is lost quickly. Correspondingly, processes at low orders in perturbation theory flip the boundary spin without costing any energy.  This behavior shows up as poles at low orders in the expansion of the almost strong zero mode  \cite{Fendley:2012}, thus obviously rendering the process invalid at this order.
For $J_2\ne 0$ the error coming from the resonances typically dominates the error described in the previous section. 

\subsection{Degeneracies in perturbation theory}

The physics resulting from $J_2\ne 0$ in (\ref{Hpert}) is different from that coming from $\Gamma_2\ne 0$. The latter is a disordering term like the transverse field, and so qualitatively does not change the physics of the ground state, as long as it is not made too large. It mainly makes its presence felt in excited-state behavior. Conversely, $J_2>0$ favors aligning next-nearest-neighbor spins, while $J_2<0$ favors anti-aligning them. The latter interaction competes with the ordering favored by the $J$ term, which no matter its sign, favors aligning spins two sites apart. Thus the physics of the ground state can and does change dramatically if $J_2<0$ and $J_2/J$ is finite. 

We will see that the physics of the highly excited states does not depend much on the sign of $J_2$. It does depend quite strongly on the {\em magnitude} of $J_2/J$, and simple perturbative arguments in the fashion of \cite{Jermyn:2014} explain why. The physics does not depend on the sign of $J$ (redefining every other spin by a spin flip maps $J\to-J$), so for simplicity we take it positive. As before, we keep $|\Gamma/J|$ and $|\Gamma_2/J|$ small, but here we allow $J_2/J$ to take on any value. Thus the disordering terms remain small relative to the ordering terms.
We write the large terms as the ``potential'' 
\begin{align}
V=-J\sum_{j=1}^{L-1}\sigma^z_{j}\sigma^z_{j+1} -J_2\sum_{j=1}^{L-2}\sigma^z_{j}\sigma^z_{j+1} \ .
\label{Vdef}
\end{align}
The eigenvalues $V_s$ of $V$ are the energies when $\Gamma=\Gamma_2=0$.

In the ordered phase, it is convenient to describe the states of the system in terms of domain walls. A domain wall occurs when two adjacent spins are different; all the states of the system can be labeled simply by specifying the domain walls, and a single spin. For $\Gamma=\Gamma_2=0$, the two states with no domain walls are the ground states $\ket{g_1}$ and $\ket{g_2}$, with energies $V_{g_1}=V_{g_2}= -J(L-1)-J_2(L-2)\equiv V_g.$
A single (or isolated) domain wall away from the edges has potential $V_g+2(J+2J_2)$. It is thus immediately apparent how frustration changes the ground states at $J_2\approx -J/2$, because creating a domain wall lowers the energy. This indicates a quantum phase transition at around this value, familiar from the studies of frustrated magnets. In general $V_s=V_g+2(n_1J+n_2J_2)$, where $n_1$ is the number of broken $J$-bonds (different nearest-neighbor spins) and $n_2$ the number of broken $J_2$ bonds (different next-nearest neighbors).

These configurations are mixed by including the single-spin-flip term with coefficient $\Gamma$. The two ground states mix first at $L$th order in perturbation theory in $\Gamma$, because it takes $L$ spin flips to get between $\ket{+++\dots}$ and $\ket{---\dots}$. Moreover, the intermediate states have at least one domain wall, and so are of potential at least $2(J+J_2)$. Thus $E_{g_1}\approx E_{g_2}$, i.e.\ the splitting between the ground-state energies is of order $(\Gamma/J)^L$.  For $J_2=0$, the same argument applies to states with domain walls. 
The resulting degeneracies are exactly those of the type arising from the Ising strong zero mode.

However, at specific nonzero values of $J_2$, this behavior changes dramatically.  Consider these two configurations, identical except for the edge spin: 
\begin{equation}
\uparrow \,\uparrow \,\downarrow\,  \cdots \qquad \Leftrightarrow \qquad  \downarrow \,\uparrow \,\downarrow\,  \cdots \label{eq:domainwallJ}\ . 
\end{equation} 
In the left configuration a single domain wall is near (but not next) to the edge; in the right the two domain walls are as close as possible to an edge. The three spins on the left contribute $2(J+J_2)$ to the potential, since it has one broken $J$-bond and one broken $J_2$-bond. On the right, the contribution is $2(2J)$, since there are two broken $J$-bonds and no broken $J_2$ bonds. Thus in the special case $J_2=J$, both configurations have the same potential. Moreover, flipping the boundary spin relates the two at first order in perturbation theory in $\Gamma$: a domain wall can be created at the edge without changing the potential. Thus at $J_2=J$ the energies of these two states, and for that matter, all one-kink states, are not exponentially close, but rather differ by a power of $1/L$.
Power-law splitting between states means that there is no pairing or partnering, so that the edge spin coherence time is not long. Indeed, we see in figure~\ref{fig:j2decay} that at $J=J_2$  the decay time is very small. The same behavior occurs for $J_2=-J$ as well, the ground states here being $\ket{--++--++\dots}$ and the flipped version. Again, a single spin flip at the boundary relates excited states and the almost strong zero mode does not occur.

This behavior is sometimes called a ``resonance'' \cite{Ros:2014}. There are many other degeneracies involving configurations with more domain walls, requiring more spin flips to go between them. Thus they will affect perturbation theory at higher orders. The almost strong zero mode provides a useful tool in finding these degeneracies, and so we describe them next.

\subsection{Poles in the strong-zero-mode expansion}
\label{sec:poles}

We have just seen that when $J_2/J$ is finite, it can be ``easy'' to flip a spin at the edge, in the sense that at low orders in perturbation theory this flip connects degenerate configurations. Here we show how the strong-zero-mode expansion provides a systematic way of understanding at which orders in perturbation theory in $\Gamma$ such behavior happens. We treat both the $J_2$ and $J$ terms as large, as opposed to section \ref{sec:aszm}. For convenience set $\Gamma_2=0$ for the most part. Including a $\Gamma_2$ term in the presence of $\Gamma\ne 0$ does not change the story qualitatively, because the double spin flip from the former appears at second-order perturbation theory in the latter. At most its presence changes the order at which a resonance appears.

The zeroth order term remains $\Psi^{(0)}=\sigma^z_1$, since it still commutes with both of the ordering terms. Commuting it with the transverse field yields 
(\ref{HPsi0}) 
as for Ising. We then find $\Gamma^{(1)}$ as in (\ref{szmrecursion}), by inverting the large terms in the Hamiltonian. Here, however, this requires inverting $[V,\cdot]$, including the next-nearest-neighbor interactions. For particular values of $J_2/J$, this may be impossible, because of extra operators in its null space. Indeed, 
\[ \Psi^{(1)}=
  \frac{\Gamma}{J^{2} - J_{2}^{2}} \sigma^x_{1}\Big(J\sigma^z_{2} - J_2\sigma^z_{3}\Big)\ .\]
The pole means the expansion collapses for $J_2=\pm J$.

This relates beautifully both approaches, illustrating that poles in the strong-zero-mode expansion correspond to processes where flipping the spins at and near the edge relates degenerate configurations \cite{Jermyn:2014}. The order in the expansion at which a pole occurs is the same order at which the corresponding easy boundary-spin flip occurs in perturbation theory. The location of the pole gives the coupling where this resonance process occurs. Moreover, the terms with a pole point to the corresponding easy boundary spin flip. For example, both terms in $\Psi^{(1)}$ contain $\sigma^x_1$, while the other two involve $\sigma^z_{2}$ and $\sigma^z_3$. This suggest we look at configurations with domain walls between spins on these sites. One is thus led very quickly to the configurations illustrated in \eqref{eq:domainwallJ} when $J_2=J$, and to the similar ones occurring when $J_2=-J$.

The nasty expressions for $\Psi^{(n)}$ can be computed with the aid of our Python program. At second order in $\Gamma$ we find
\begin{widetext}
\begin{align*}
  \Psi^{(2)}&=\frac{J J_{2} \Gamma^{2}}{(J^2-J_2^2)^2} \sigma^z_{1} \sigma^z_{2} \sigma^z_{3}+ \frac{ \Gamma^2}{(J^2-J_2^2)^2(9J^2- J_2^2)} \sigma^x_{1} \sigma^x_{3} \Big[2 J^2 J_2^2 \sigma^z_{2} \sigma^z_{4} \sigma^z_{5} + 2 J^2 J_2^2 \sigma^z_{1} \sigma^z_{3} \sigma^z_{5} - 6 J^3 J_2 \sigma^z_{1}\sigma^z_{3} \sigma^z_{4}
\\
&  \quad + 2 J^2 J_2^2 \sigma^z_{1} \sigma^z_{2} \sigma^z_{3} \sigma^z_{4} \sigma^z_{5} - J J_2\left(J_{2}^{2}- 3 J^{2}\right)(\sigma^z_2+ \sigma^z_{4}-\sigma^z_{1} \sigma^z_{2} \sigma^z_{3}) + J_2^2\left(J_{2}^{2}- 7 J^2 \right) \sigma^z_{5} \Big]\\
&\quad +\frac{\Gamma^2}{(J^2-J_2^2)(J^2-4 J_2^2)} \sigma^x_{1} \sigma^x_{2}\left[-J J_{2} \sigma^z_{1} \sigma^z_{2} \sigma^z_{3} - J J_{2}  \sigma^z_{4}  + 2 J_{2}^{2} \sigma^z_{1} \sigma^z_{2} \sigma^z_{4} + \left(J^{2} - 2 J_{2}^{2}\right) \sigma^z_{3} \right] \ .
\end{align*}
\end{widetext}
New poles at $J_2 = \pm \frac{J}{2}$ and $J_2 = \pm 3J$ are immediately apparent.  The poles in the considerably nastier expression for $\Psi^{(3)}$ are at $J_2=J/3$, $J_2=3J/2$ and $J_2 = 5 J$. 
It is not difficult to find the corresponding processes that flip the boundary spin without changing the energy. The terms in $\Psi^{(2)}$ with a pole at $J_2 = 3J$ contain spin-flip terms $\sigma^x_{1}\sigma^x_3$, and otherwise sample up to the fifth lattice site. This indicates the configurations that have the same energy when $J_2=3J$ are related by flipping the first and third spins. They are
\begin{align}
\uparrow \,\uparrow \,\uparrow \,\uparrow \,\downarrow\, \cdots \quad  \Leftrightarrow \quad  \downarrow \,\uparrow \,\downarrow\, \uparrow\,  \downarrow\, \cdots \qquad J_2 = 3 J\ , \label{eq:domainwall3J}
\end{align}
Likewise, the two configurations
\begin{align}
 \uparrow \,\downarrow \,\downarrow \,\uparrow \, \cdots \quad & \Leftrightarrow \quad  \downarrow \,\uparrow \,\downarrow\,  \uparrow\, \cdots \qquad J_2 = \frac{J}{2}\ .
 \label{eq:domainwallhalfJ}
\end{align}
have potentials $2(2J+2J_2)$ and  $2(3J)$ respectively, equal at $J_2=J/2$. They are related by flipping the first two spins, with the corresponding pole terms in $\Psi^{(2)}$ indeed containing 
$\sigma^x_1\sigma^x_2$. It is worth noting that when $\Gamma_2\ne 0$, this process occurs at first order, so that a term with coefficient proportional to $\Gamma_2/(J-2J_2)$ appears in $\Psi^{(1)}$.

The above arguments readily generalize to find other degenerate states related by spin flips at and near the edge. For example, for $J_2=rJ$ for $r$ any odd integer, we have:
\[\underbrace{\uparrow\,\uparrow\,\uparrow\,\uparrow\,\cdots\,\uparrow\,\uparrow}_{r+1\text{ up spins}}\downarrow\,\cdots
 \quad\Leftrightarrow\quad \underbrace{\downarrow\,\uparrow\,\downarrow\,\uparrow\,\cdots\,\downarrow\,\uparrow}_{r+1\text{ alternating spins}}\downarrow\,\cdots
 \]
 These are related by the spin flips $\sigma^x_{1}\sigma^x_3\cdots\sigma^x_{r}$.
This process creates $r$ domain walls (broken $J$ bonds), while decreasing the number of broken $J_2$ bonds by one by making the spins at sites $j=r$ and $r+2$ the same. The resulting change is potential is $\Delta V=2(rJ-J_2)$, which indeed vanishes when $J_2=rJ$.

The poles for $J_2=J/r$ for $r$ an integer are interesting, since they enable a conjecture for asymptotic formula for the decay time valid when $J_2/J$ is small \cite{KLY}. First consider a configuration with the maximum number of broken $J_2$ bonds within the first $3(r+1)/2$ spins:
\[\underbrace{\uparrow\,\uparrow\,\downarrow\,\downarrow\,\uparrow\,\uparrow\,
\downarrow\,\downarrow\,\cdots}_{3(r+1)/2\text{ spins}}
 \]
For $r$ odd, act on this with the flips $\sigma^x_1\sigma^x_4\cdots\sigma^x_{(3r-1)/2}$. This decreases the number of broken $J_2$ bonds by $r$, because flipping the first spin heals one $J_2$ bond, while the remaining $(r-1)/2$ flips heal two bonds each. This process increases the number of broken $J$ bonds (i.e.\ the number of domain walls) by one, since flipping the first spin creates one domain wall, while the remaining flips do not change the number, but rather just move a domain wall. Thus for this process the change in potential is $\Delta V =2(J-rJ_2)$, indeed giving a resonance at $J_2=J/r$ for $r$ an odd integer. Similarly, for $r$ an even integer, consider the configuration 
\[\uparrow\underbrace{\downarrow\,\downarrow\,\uparrow\,\uparrow\,
\downarrow\,\downarrow\,\uparrow\,\uparrow\cdots}_{3r/2\text{ spins}}\]
which also has the maximum number of broken $J_2$ bonds. Here the combination of flips $\sigma^x_1\sigma^x_2\sigma^x_5\sigma^x_8\cdots\sigma^x_{3(r-2)/2}$ gives the same $\Delta V =2(J-rJ_2)$. Thus there are poles in the zero mode expansion at $J_2=J/r$ for all $r$.

These are not the unique processes yielding poles at $J_2=J/r$. The preceding process for $r=3$ occurs only in $\Psi^{(4)}$, whereas the pole at $J_2=J/3$ in $\Psi^{(3)}$ involves $\sigma^x_1\sigma^x_3\sigma^x_4$. However, degeneracies giving resonances at $J_2=J/r$ occur only for states whose energy grows with $r$ as $rJ$. The reason is that one needs at least $r$ broken $J_2$ bonds in one of the states to have this degeneracy, and this implies at least $r/2$ broken $J$ bonds in that state. Thus the splitting due to the higher order poles occurring at $J_2/J$ small only should affect highly excited states, i.e.\ those whose energy density relative to the ground state is nonzero. This is in accord with our argument that for small $J_2$ the pairing with exponentially small splitting should still persist for eigenstates with zero energy density.

\subsection{The effect of the resonances}

The resonances and the ensuing poles are a generic property for Hamiltonians with two different ordering terms. Given the plethora of processes occurring for a large number of broken bonds near the boundary, it is natural to expect that the poles occur at all rational values of $J_2/J$. For most values, however, the poles will presumably occur at very high orders in perturbation theory, and the arguments in section \ref{sec:aszm} show that the expansion must truncated at some finite $n_*$ already. Thus one might hope that the poles can be handled by truncating the expansion before the pole occurs. A
worry though is that even if one tunes $J_2$ away from a pole (say to $J_2/J$ irrational), nearby poles will make the norm of the error term large, and destroy the edge coherence quickly. 
We use exact diagonalization here to indicate that this worry is unfounded. In fact, we find that  when $\Gamma_2=0$, the edge spin coherence time is quite large as long as $J_2$ is not very close to a low-order pole. 

\begin{figure} [htbp]
\centering
 \includegraphics[width=\linewidth]{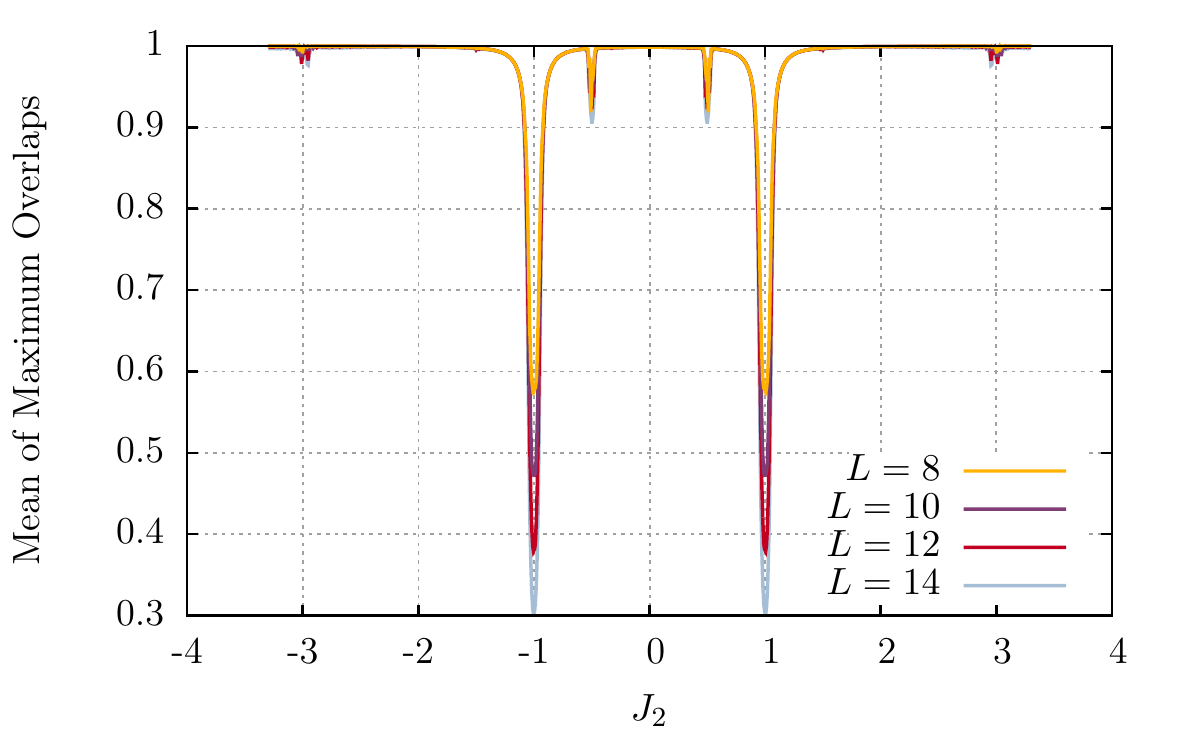}
\caption{The mean of the maximum overlaps between eigenstate pairs and $\sigma^z_1$ for $\Gamma=0.05$. 
The resonances are clearly visible.}
\label{fig:bmj2}
\end{figure}

We first check that while the resonances drastically decrease the overlap at the poles, at other values of the couplings the overlap remains large at the sizes we can access with exact diagonalization. To make Figure~\ref{fig:bmj2}, we compute the overlap $\Braket{s|\sigma^z_1 |s'}$ for all $s,s'$, find the maximum magnitude for each $s$, and then average this value over all $s$, that is:
 \[\frac{1}{2^{L}}\sum_s\left[ \max_{s^\prime} \Braket{s|\sigma^z_1|s^\prime}
\right]\]  At the small value $\Gamma=0.05$ the pairing away from the poles is almost perfect, and that the decrease in pairing even for the second order pole with system size is very slow. Presumably at larger system sizes the effects of more poles will appear, but in between poles the overlap will still be substantial. Note that this graph is also symmetric about $J_2 = 0$, indicating that this structure is independent of the different ground-state physics occurring when $J_2$ is antiferromagnetic.
This plot is a strong sign that the resonances are the dominant effect for non-vanishing $J_2$.

If the paired states shown in figure \ref{fig:bmj2} had exactly the same energy, the edge coherence time would survive as long as the pairing does. However, it does not. Recall that in figure~\ref{fig:Ising_log_decay} we observed `revivals' in $A_\infty(t)$ at long time scales because all the energy difference between paired states were the same (due to a quirk of free fermions). This illustrates clearly the idea that it is not the energy differences of paired states that cause a decay to zero in  $A_\infty(t)$, but rather their \emph{variance}, as it is this variance which drives dephasing. 
In order to investigate more closely the effects of the resonance on the coherence time of the edge spin, we introduce the paired energy-difference variance (PEDV). 
For each pair of eigenstates $s,s'$ defined by maximum overlap using $\sigma^z_1$, we find the energy difference $\Delta_s\equiv E_s-E_{s'}$. The PEDV is the variance of all these energy differences, where the average is over all eigenstates.  
As well as indicating how consistently an almost strong zero mode starting with  $\sigma^z_1$ pairs the energy spectrum, the inverse of the PEDV gives an estimate of the timescale on which the decay from the plateau at $\mathcal{N}^2$ occurs. 


\begin{figure} [htbp]
\centering
 \includegraphics[width=\linewidth]{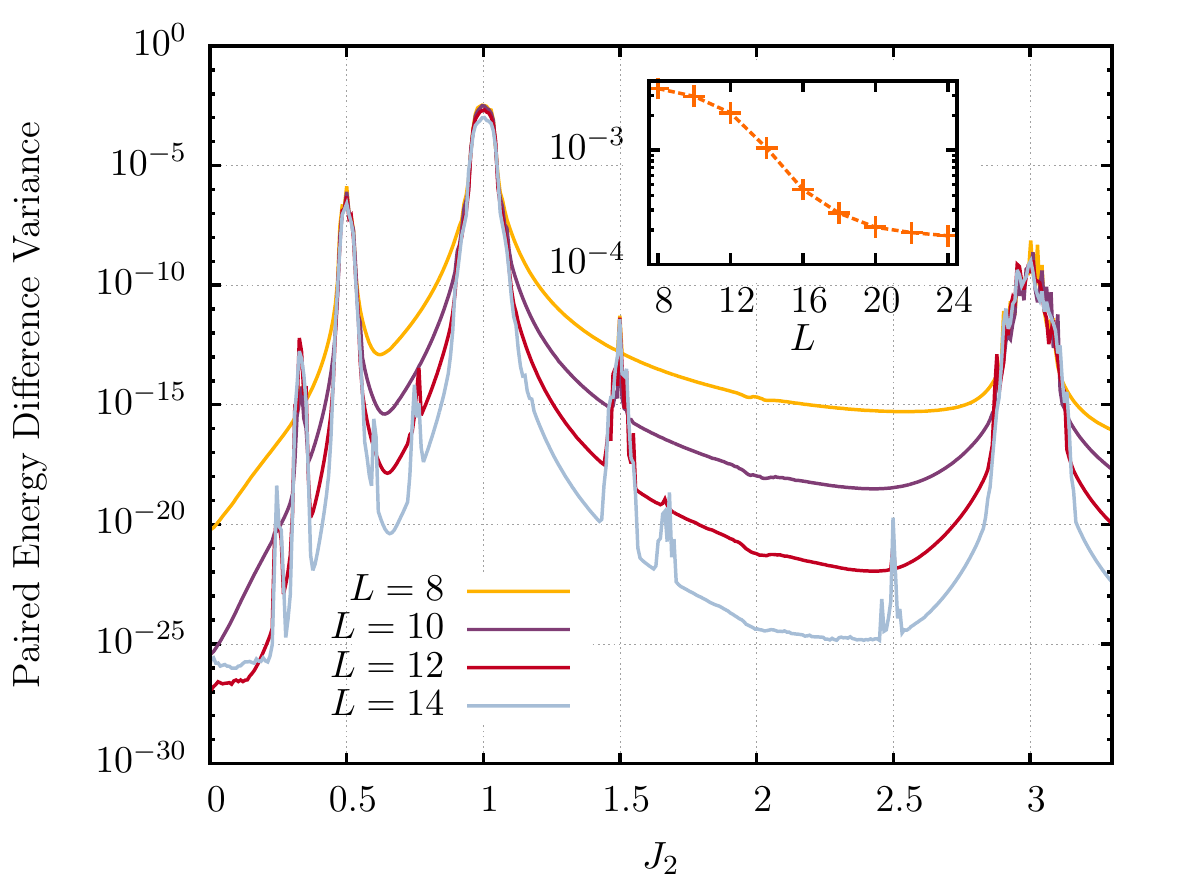}
\caption{The PEDV (see text) calculated for $\Gamma =0.05$ using exact diagonalisation. Inset: The height of the first order peak in PEDV at $J_2 = J$ plotted against system size.}
\label{fig:pedvj2}
\end{figure}
The PEDV is a much more sensitive probe of the resonances than the pairing, and is plotted for the $J_2$ large case in Figure~\ref{fig:pedvj2}.
The peaks in the energy differences measured by the PEDV occur precisely at the couplings with poles in the strong zero mode expansion. Not only are the first- and second-order poles visible but so are the third-order poles at $J = J/3$, $3J/2$ and at the larger system sizes even fourth-order poles at $J_2 = J/4$ and $3 J/4$. 
Away from the peaks/poles the PEDV exponentially decreases with system size. Notice the log scale on the y axis: this structure in the PEDV traverses almost 30 orders of magnitude! This is the finite-size behavior of an exact strong zero mode, as we checked by computing the PEDV for XYZ. Of course, we do not expect this exponential decrease to survive in the $L\to\infty$ limit, where presumably {all} points lie arbitrarily close to a pole.

\begin{figure} [htbp]
\centering
 \includegraphics[width=\linewidth]{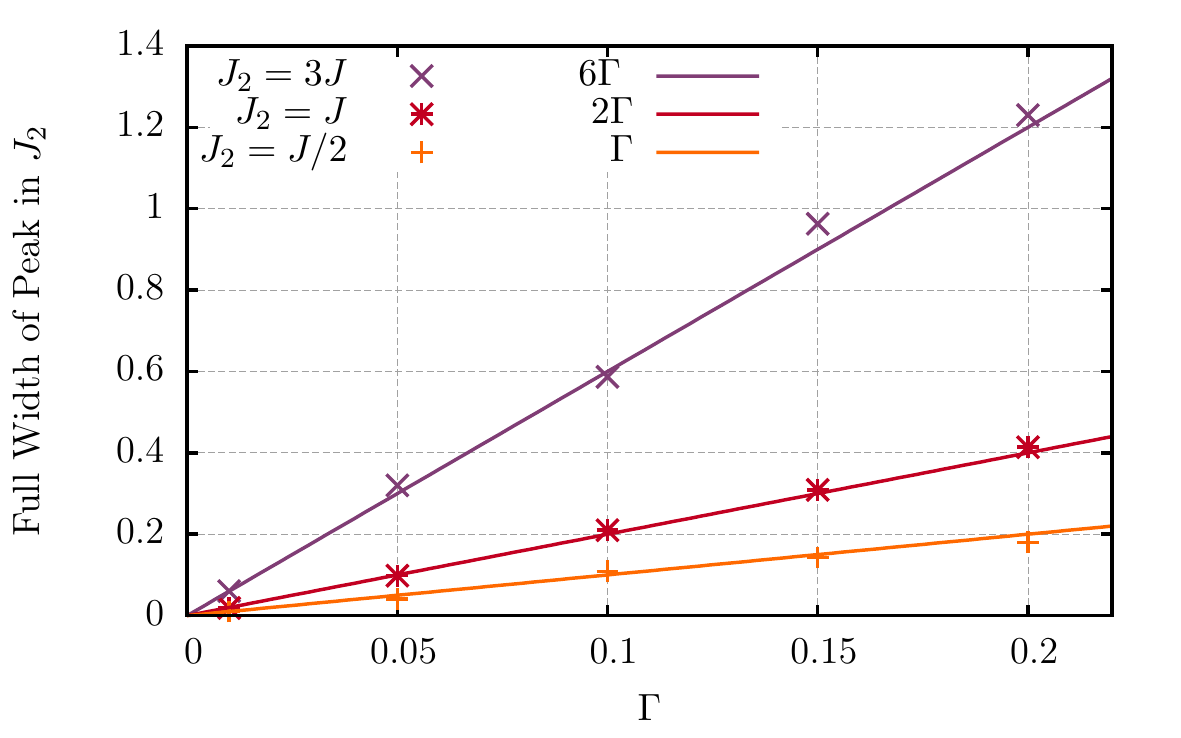}
\caption{The width of the first- and second-order poles in the PEDV as a function of $\Gamma$ from exact diagonalisation (points) and the theoretical prediction $2\Gamma{J_2}/{J} $  (lines). }
\label{fig:width}
\end{figure}

On the other hand, close to the poles the PEDV appears to saturate with $L$. Importantly, the peaks are {\em not} increasing in height or width.  The width of the peak affects how far away from a pole we must tune the couplings in order to avoid its effect. It is clear from figure~\ref{fig:pedvj2} that the width converges with $L$ -- all the curves of different $L$ intersect when they transition from finite-size behaviour (exponentially decreasing with $L$) to resonant behaviour. Moreover, the width depends on the couplings as
\[\hbox{width } \sim 2\frac{J_2}{J} \Gamma\]
as is clear from figure \ref{fig:width}.

This behaviour may be explained heuristically by the following argument.  The resonances are caused by easy edge-spin-flip processes that convert between between states with the same potential by exchanging $J$ and $J_2$ bonds. When we plot the PEDV against $J_2$, we are implicitly testing edge-spin flip processes where we know the energy of the $J_2$ bonds we are sending in, but are affected by the energy uncertainty of produced $J$ bonds. Moreover, for non-zero $\Gamma$ domain walls have a finite lifetime and therefore an energy uncertainty $\sim \Gamma$. So for example close to the $J_2 = 3 J$ pole, the flip converts one $J_2$ bond to three $J$ bonds, each with associated energy uncertainty $\Gamma$. We thus expect the half-width of the pole in $J_2$ to be $3 \Gamma$. Likewise, at $J_2 = J/2$ we convert two $J_2$ bonds to one $J$ bond, so the half-width is reduced to $\Gamma/2$. In general we thus expect the half-width to be $\Gamma{J_2}/{J}$, as the data indicate.

At large values of $\Gamma$ the peaks widen and begin to merge, so the estimate for the width of the PEDV peaks breaks down.  For the $J_2=0$ case, the strong zero mode seems to go away at roughly the value of $\Gamma$ where the phase transition to a disordered phase occurs. The same is likely true here as well, as long as a resonance has not already killed it.

The fact that the PEDV curves for different $L$ intersect suggest that the peak height also saturates with $L$.
A closer look at the first order pole at $J_2 = J$ in the inset of Figure~\ref{fig:pedvj2} reveals the log scale is hiding a substantial (but not exponential) decrease in the peak of the pole with $L$. However, at larger system sizes, saturation of peak height does appear to occur. This implies that the decay time of the autocorrelation of the edge spin $A_\infty(t)$ will also saturate with $L$ at and near the poles, rather than increasing exponentially (as it would if there were an exact strong zero mode), or decreasing (as it would if there were not even an almost strong zero mode).

It is important to stress that we have have discussed pairing and partnering stemming from operators at the edge. However, pairing in the energy spectrum could result from another, not edge-localised (or even local) operator. For example, at the pole at $J_2 = J$ it turns out that although the pairing due to $\sigma^z_1$ dies, the pairing due to $\sigma^z_2$ is large. Indeed, if one starts the strong zero mode expansion with $\sigma^z_2$, it is even more complicated, but it does lack the pole at $J_2 = J$. This is simple to understand by referring back to \eqref{eq:domainwallJ} and noting that there are no easy `edge' spin-flip processes that flip $\sigma^z_2$. Thus our results do not preclude long coherence times for other quantities when $J_2=J$.

One final note: the PEDVs here are calculated for small $\Gamma$ and large $J_2$, which imply the existence of two very different scales. Widely separated scales can lead to spurious signs of localization in small-size numerics 
coming from `minibands', gaps in the density of states \cite{papic_many-body_2015}. We also observe minibands in our numerics. These are most prominent when $J_2$ is a rational (or even more so, an integer) multiple of $J$, as this reduces the number of minibands possible. Nevertheless, we are confident that the physics we describe is independent of these minibands. One reason is that minibands occur at any rational $J_2/J$, whereas our phenomena depend strongly on which particular rational value. Another is that
when we replace the $\Gamma$ term by the $\Gamma_2$ term and repeat the calculation, we find the poles change in position and relative magnitude exactly as predicted by the strong-zero-mode expansion. This behavior is not explained by any miniband structure, which should not depend on which disordering term is used, but only the magnitude.

\section{Conclusion}

We have demonstrated that at the edge of certain spin chains, quantum coherence is preserved for long times. This holds because of, not in spite of, strong interactions between the spins. The ground state must be ordered, but the long coherence times occur even if the initial state is at infinite temperature. Although in some ways the physics resembles that arising in many-body-localization, the systems we analyze have no disorder.

This behavior arises because of the presence of an almost strong zero mode. This operator is localized at the edge, and almost commutes with the Hamiltonian. When spins are mapped to fermions via the Jordan-Wigner transformation, order becomes topological order, and an edge zero mode guarantees the ground-state degeneracy. The consequences of a strong zero mode are even more dramatic: it implies an intimate relation between states in different symmetry sectors, and this relation underlies the long coherence time.

Many interesting directions remain to be explored. The unusual properties of the strong zero mode first became apparent in studies of parafermionic systems \cite{Fendley:2012,Jermyn:2014}, but it is still not clear even in the integrable cases whether or not the strong zero mode is in general almost or not. Certainly resonances appear, and this and other properties have been analyzed \cite{Moran:2017}. 

The system studied in \cite{Yao:2015} is similar to the perturbed Ising model studied in this paper, in that it not integrable and can be expressed in terms of Majorana fermions with four-fermion interactions. That of \cite{Yao:2015} is particularly interesting in that there are potentially two zero modes at each edge, so that there can be a qubit, a two-state system at each edge. Many of the same considerations, in particular the destruction of coherence by the resonances at some couplings, apply. These will be analyzed in \cite{KLY}.

At a surface level, the results here clearly resemble those coming from prethermalization, in that certain physical quantities take a long time to reach their equilibrium state. At a deeper level, both types of phenomena have at their heart an almost conservation law or laws. Understanding generally how the strong zero mode relates to integrability is very much still an open problem, but the connection to prethermalization is likely to shed a great deal of light on the situation. This connection can be made precise by utilizing the rigorous approach of \cite{Abanin} to prethermalization, and these results will be detailed in \cite{EFKN}.

\bigskip

\paragraph{Acknowledgments} We thank F.\ Essler, M.\ M\"uller, and C.\ Nayak for very helpful conversations.  
This work of P.F.\ was supported by EPSRC through grant EP/N01930X, and N.Y.\ 
by the Miller Institute for Basic Research in Science. C.R.L.\ acknowledges support from the Sloan Foundation through a Sloan Research Fellowship and the NSF through grant PHY-1656234.

\bibliography{szm}
\bibliographystyle{apsrev4-1}

\end{document}